\documentclass[
 amsmath,amssymb,
prb,
twocolumn,
]{revtex4-2}
\usepackage{dcolumn}
\usepackage{bm}
\usepackage{braket}
\usepackage{color}
\usepackage{accents}
\usepackage{mathrsfs}
\usepackage{bm}
\usepackage{here}
\usepackage{longtable}

\usepackage{mathtools}
\usepackage{empheq}
\usepackage{graphicx}
\usepackage{hyperref}
\usepackage{comment}

\definecolor{dgreen}{RGB}{00, 120, 00} \definecolor{dblue}{RGB}{00, 00, 220}
\definecolor{lgreen}{RGB}{46, 139, 87}
\usepackage{cancel}
\usepackage{ulem}

\newcommand{\add}[1]{\textcolor{blue}{#1}}

\begin{document}
\title{
Discontinuous Transition to Superconducting Phase
}
\author{Takumi Sato$^{1}$}
\author{Shingo Kobayashi$^{2}$}
\author{Yasuhiro Asano$^{1}$}%
\affiliation{
$^{1}$Department of Applied Physics, Hokkaido University, Sapporo 060-8628, Japan.\\
$^{2}$RIKEN Center for Emergent Matter Science, Wako, Saitama 351-0198, Japan\\
}

\date{\today}

\begin{abstract}
We discuss the instability of uniform superconducting states
that contain the pairing correlations belonging to the odd-frequency symmetry class.
The instability originates from the paramagnetic response
of odd-frequency Cooper pairs and is considerable at finite temperatures.
As a result, the pair potential varies discontinuously at the transition temperature
when the amplitude of the odd-frequency pairing correlation functions 
is sufficiently large.
We also discuss the universality of the phenomenon.
\end{abstract}
\maketitle

\section{Introduction}
There are two types of uniform perturbations that act on uniform superconducting states.
One does not change the thermal properties, while the other does.
Spin-orbit interactions and Zeeman fields correspond to the examples of such perturbations 
in a spin-singlet superconductor.
Spin-orbit interactions do not change any thermal properties of a superconductor
such as the transition temperature $T_c$ or the dependence of order parameter 
$\Delta$ on temperatures $T$~\cite{frigeri:prl2004}.
On the other hand, uniform Zeeman fields decrease $T_c$.
Moreover, the transition to the superconducting phase by decreasing the temperature
changes to a first-order transition in sufficiently strong Zeeman fields~\cite{sarma:jpcs1963,maki_tsuneto:ptp1964}.
Namely, the superconducting state is thermally unstable under Zeeman fields.
A recent study~\cite{menke:prb2019} has reported that
$j=3/2$ superconductors also exhibit very similar instabilities.
Although such discontinuous transition has been observed in spin-singlet superconductors
under Zeeman fields~\cite{bianchi:prl2002,radovan:nature2003,lortz:prl2007},
there has been no comprehensive explanation for why the 
superconducting transition changes to a first-order transition and 
what distinguishes the two types of perturbations.
We address these issues in the present paper.

To the best of our knowledge,
odd-frequency Cooper pairs~\cite{berezinskii:jetplett1974,bergeret:rmp2005,tanaka:jpsj2012,linder:rmp2019,triola:annphys2020,cayao:epj2020}
tend to cause 
thermal instability in superconducting states.
This consideration is supported by the following findings 
for odd-frequency pairs localized various places in a superconductor such as 
at a vortex core~\cite{tanuma:prl2009}, 
in the vicinity of a magnetic cluster~\cite{kuzmanovski:prb2020,perrin:prl2020},
and at the surface of a topologically nontrivial superconductor~\cite{tanaka:prl2007,asano:prb2013}.
An analysis of the free-energy density shows that the superconducting states 
are unstable locally around these defects~\cite{suzuki:prb2014,shu:prb2022}.
The paramagnetic response of odd-frequency Cooper pairs to an external magnetic field
is responsible for the instability~\cite{asano:prl2011}.
In uniform superconductors, odd-frequency Cooper pairs exist as subdominant pairing correlations
when the electronic structures have extra degrees of freedom such as 
spins, orbitals and sublattices~\cite{BSchaffer:prb2013}.
It has been shown that the $T_c$ of such superconductors decreases as
the amplitude of odd-frequency pairs increases~\cite{asano:prb2015}.

The purpose of this paper is to show that odd-frequency Cooper pairs in uniform
superconducting states
are responsible for the discontinuous transition to the superconducting phase.
For this purpose,
we analyze the way in which the odd-frequency pairing correlation functions
change the coefficient of the $\Delta^4$ term in the Ginzburg-Landau (GL)
free-energy functional and the superfluid density.
We find that the odd-frequency pairing correlations decrease
the coefficient and the superfluid density in the same manner.
The instability originates from
the suppression of the superfluid density due to odd-frequency pairs.
    We conclude that the discontinuous transition to the superconducting phase
    is a common feature of superconductors that contain a large amount of
    uniform paramagnetic odd-frequency Cooper pairs in their superconducting phase.
The two types of uniform perturbations are distinguished by whether or not they
induce odd-frequency Cooper pairs.

This paper is organized as follows.
In Sec.~\ref{sec:GL}, we explain a model of $j=3/2$ superconductors
which we mainly analyze in this paper
and show the expression of the coefficients in the GL free-energy functional
in terms of the Green's function.
The discontinuous transition to the superconducting phase
is demonstrated numerically in Sec.~\ref{sec:discontinuous}.
The mechanism of the discontinuous transition is discussed
by analyzing the temperature dependence of the superfluid density in Sec.~\ref{sec:pairdens}.
In Sec.~\ref{sec:zeeman}, we discuss the universality of the phenomenon
by showing the discontinuous transition in a spin-singlet superconductor in Zeeman fields
and that in a two-band superconductor under the band-hybridization.
The conclusions are given in Sec.~\ref{sec:conclusion}.

\section{Ginzburg-Landau Free-energy}
\label{sec:GL}

\subsection{Multi-band superconductors}
In this paper, we mainly analyze the Hamiltonian of pseudospin-quintet states in a $j=3/2$ 
superconductor for the following several reasons.
The normal state Hamiltonian describes the most general multiband electronic states,
which have four internal degrees of freedom and preserve
both time-reversal symmetry and inversion symmetry~\cite{cavanagh:prb2023}.
The pair potential can be represented by a simple formula~\cite{agterberg:prl2017,brydon:prb2018}.
Useful mathematical tools are available to calculate the Green's function analytically. 
The high-pseudospin electronic states stem from the strong coupling between orbitals with angular 
momentum $\ell=1$ and spin with $s=1/2$~\cite{luttinger:pr1955,brydon:prl2016,agterberg:prl2017,brydon:prb2018,roy:prb2019}.
The mean-field Hamiltonian can be expressed as
\begin{align}
    \label{eq:hbcs}
    \mathcal{H} &=
    \frac{1}{2} \sum_{\bm{k}}
    \vec{\Psi}^{\dag}_{\bm{k}} H(\bm{k}) \vec{\Psi}_{\bm{k}}
    + \frac{N\Delta^2}{\tilde{g}} , \\
    \vec{\Psi}_{\bm{k}} &=
    \left[
        \vec{\psi}^{\,\mathrm{T}}_{\bm{k}}, \, \vec{\psi}^{\,\dag}_{-\bm{k}}
    \right]^{\mathrm{T}} ,\\
	\vec{\psi}_{\bm{k}} &=
    \left[
        c_{\bm{k},3/2}, \, c_{\bm{k},1/2}, \,
        c_{\bm{k},-1/2}, \, c_{\bm{k},-3/2}
    \right]^{\mathrm{T}}, 
\end{align}
where $\tilde{g}>0$ represents the strength of the attractive interaction,
$N$ is the number of unit cells of the underlying lattice,
$\Delta$ denotes the pair potential, 
and $c_{\bm{k},j_z}$ is the annihilation operator of an electron at $\bm{k}$ with pseudospin 
$j_z$. 
The Bogoliubov-de Gennes (BdG) Hamiltonian in Eq.~\eqref{eq:hbcs} is,
\begin{align}
    \label{eq:Hbdg_j3/2}
    H(\bm{k}) &=
    \left[
        \begin{array}{cc}
            H_{\mathrm{N}}(\bm{k}) & \Delta(\bm{k}) \\
            -\Delta^\ast(-\bm{k}) & -H^\ast_{\mathrm{N}}(-\bm{k})
        \end{array}
    \right] .
\end{align}
The normal state Hamiltonian is represented by the tight-binding model on a simple cubic lattice~\cite{luttinger:pr1955} as
\begin{align}
    H_{\mathrm{N}} (\bm{k}) &=
    -2t_1 \sum_{\nu} \cos{k_{\nu}}
    -2t_2 \sum_{\nu} \cos{k_{\nu} } \, J_{\nu}^2 \nonumber \\
    &+ 4t_3 \sum_{\nu \neq \nu'} \sin{k_{\nu}} \sin{k_{\nu'}} \, J_{\nu} J_{\nu'} + 6 t_1 + \frac{15}{2} t_2 - \mu , \nonumber \\
    &=
    \xi_{\bm{k}} + \vec{\epsilon}_{\bm{k}} \cdot \vec{\gamma} ,
    \label{eq:hn}
\end{align}
with $\mu$ being the chemical potential.
The corresponding point group is $O_h$.
The nearest neighbor hopping independent of (depending
on) pseudospin is $t_1$ ($t_2$). 
The second neighbor hopping is denoted by $t_3$.
$\xi_{\bm{k}}$ represents kinetic energy of an electron and
the five-component vector
$\vec{\epsilon}_{\bm{k},\mathrm{j}}$ for $\mathrm{j}=1-5$ determines
the dependence of the normal state dispersions on pseudospins.
The expressions of $\vec{\epsilon}_{\bm{k},\mathrm{j}}$ and 
five $4\times 4$ matrices $\gamma^{\mathrm{j}}$ for $\mathrm{j}=1-5$ are given in Appendix \ref{sec:algebras}.
The pair potential is represented by
%
%
\begin{align}
    \Delta(\bm{k}) &= \Delta \, \vec{\eta}_{\bm{k}} \cdot \vec{\gamma} \, U_T , \label{eq:def_pairpotential}\\
    \vec{\eta}_{\bm{k}} &= (\eta_{\bm{k},1},\eta_{\bm{k},2},\eta_{\bm{k},3},\eta_{\bm{k},4},\eta_{\bm{k},5}) ,
	\label{eq:eta_vec0}
\end{align}
where the five-component vector $\vec{\eta}_{\bm{k}}$
with $|\vec{\eta}_{\bm{k}}| = \sqrt{ \vec{\eta}_{\bm{k}} \cdot \vec{\eta}^{\,\ast}_{\bm{k}} } = 1$
represents an even-parity pseudospin-quintet pairing order.
The Fermi-Dirac statistics of electrons implies
\begin{align}
\Delta^{\mathrm{T}}(-\bm{k}) = - \Delta(\bm{k}).
\end{align}
Here $\mathrm{T}$ means the transpose of the matrix which represents the interchange
of pseudospins at two electrons in a Cooper pair.
Since $ [ \vec{\eta}_{\bm{k}} \cdot \vec{\gamma} \, U_T ]^\mathrm{T}= -  \vec{\eta}_{\bm{k}} \cdot \vec{\gamma} \, U_T $,
the pseudospin-quintet states are antisymmetric under interchanging two pseudospins.

%
%


\begin{widetext}
\subsection{Ginzburg-Landau expansion}
\label{subsec:gl_expansion}
To analyze superconducting states, we solve the Gor'kov equation
\begin{align}
    \left[ i\omega_n - H(\bm{k}) \right] 
    \left[
        \begin{array}{rr}
            \mathcal{G}(\bm{k},i\omega_n) & \mathcal{F}(\bm{k},i\omega_n) \\
            -\undertilde{\mathcal{F}}(\bm{k},i\omega_n) & -\undertilde{\mathcal{G}}(\bm{k},i\omega_n)
        \end{array}
    \right] =1,
    \label{eq:gorkov}
\end{align}
where $\omega_n = (2n+1) \pi T$ is the fermionic Matsubara frequency with $n$ being an integer,
and $\undertilde{X}(\bm{k}, i\omega_n) \equiv X^{\ast}(-\bm{k}, i\omega_n)$ represents the particle-hole conjugation of $X(\bm{k}, i\omega_n)$.
%
%
The anomalous Green's function satisfies
$\mathcal{F}^{\mathrm{T}} (-\bm{k},-i\omega_n) = -\mathcal{F} (\bm{k},i\omega_n)$
due to the Fermi-Dirac statistics of electrons.
The GL free-energy functional per unit cell
is represented in terms of the Green's function~\cite{eilenberger:zphys1966,suh:prr2020}
\begin{align}
    \Omega_{\mathrm{SN}}(\Delta) &=
    \label{eq:gl_func}
    a \Delta^2 + b \Delta^4 + c \Delta^6 + \text{h. o. t.} , \\
    a \Delta^2 &=
    \frac{\Delta^2}{\tilde{g}} + T\sum_{\omega_n} \frac{1}{N}\sum_{\bm{k}}
    \frac{1}{2} \mathrm{Tr}\left[ \mathcal{F}_1 (\bm{k},i\omega_n) \Delta^{\dag}(\bm{k}) \right] , \label{eq:adef_gl} \\
    b \Delta^4 &= T\sum_{\omega_n} \frac{1}{N}\sum_{\bm{k}}
    \frac{1}{4} \mathrm{Tr}\left[
    \mathcal{F}_1 (\bm{k},i\omega_n) \Delta^{\dag}(\bm{k})
    \mathcal{F}_1 (\bm{k},i\omega_n) \Delta^{\dag}(\bm{k}) \right] , \label{eq:bdef_gl} 
\end{align}
where $\mathcal{F}_1 (\bm{k},i\omega_n) \equiv -\mathcal{G}_{\mathrm{N}}(\bm{k},i\omega_n)\, \Delta(\bm{k}) \,
                                           \undertilde{\mathcal{G}}_{\mathrm{N}}(\bm{k},i\omega_n)$
is the anomalous Green's function within the first order of $\Delta$ and  
$\mathcal{G}_{\mathrm{N}}
(\bm{k},i\omega_n)=\left[ i\omega_n - H_{\mathrm{N}}(\bm{k}) \right]^{-1}$ is the Green's function in the normal state.
%
In a usual second-order transition, the equation $a=0$ gives the transition temperature $T_c$.
The inequalities $a<0$ and $b>0$ describe the stable superconducting state for $T<T_c$.

The anomalous Green's function for Eqs.~\eqref{eq:hn} and \eqref{eq:def_pairpotential} consists of four components as
\begin{align}
    \mathcal{F}_1 (\bm{k},i\omega_n)
    \label{eq:f1}
    &= \frac{\Delta}{Z_0}
    \left[
        f_1^{\Delta}(\bm{k},i\omega_n) + f_1^{\mathrm{s}}(\bm{k},i\omega_n) + f_1^{\mathrm{q}}(\bm{k},i\omega_n) + f_1^{\mathrm{odd}}(\bm{k},i\omega_n)
    \right] , \\
    f_1^{\Delta}(\bm{k}, i\omega_n) &= -( \omega_n^2 + \xi_{\bm{k}}^2 ) \vec{\eta}_{\bm{k}} \cdot \vec{\gamma} \, U_T , 
	\quad
    f_1^{\mathrm{s}}(\bm{k}, i\omega_n) = 2\xi_{\bm{k}} \, \vec{\eta}_{\bm{k}} \cdot \vec{\epsilon}_{\bm{k}} \, U_T , \quad
    f_1^{\mathrm{q}}(\bm{k}, i\omega_n) = - \vec{\epsilon}_{\bm{k}} \cdot \vec{\gamma} \, \vec{\eta}_{\bm{k}} \cdot \vec{\gamma} \, \vec{\epsilon}_{\bm{k}} \cdot \vec{\gamma} \, U_T , \\
    f_1^{\mathrm{odd}}(\bm{k}, i\omega_n) &= -i\omega_n P_{\mathrm{O}} \, U_T , \quad
    P_{\mathrm{O}} =
    \label{eq:Po_j32}
    \left[ \vec{\eta}_{\bm{k}} \cdot \vec{\gamma} \, , \, \vec{\epsilon}_{\bm{k}} \cdot \vec{\gamma} \right], \\
    Z_0 &= ( \omega_n^2 + \xi_{\bm{k}}^2 - \vec{\epsilon}_{\bm{k}}^{\,2} )^2 + 4 \omega_n^2 \vec{\epsilon}_{\bm{k}}^{\,2} 
    = \xi_{\bm{k}}^4 + 2 \xi_{\bm{k}}^2 (\omega_n^2 - \vec{\epsilon}_{\bm{k}}^{\,2}) 
    + (\omega_n^2+ \vec{\epsilon}_{\bm{k}}^{\,2})^2 ,
\end{align}
%
%
with $\left[ A \, , \, B \right] = A B - B A$.
$f_1^{\Delta}$ in Eq.~\eqref{eq:f1} belongs to pseudospin-quintet symmetry and is linked to the pair potential through the gap equation.
The spin-orbit interactions $\vec{\epsilon}_{\bm{k}}$ induce a pseudospin-singlet correlation function $f_1^{\mathrm{s}}$
and another pseudospin-quintet correlation function $f_1^{\mathrm{q}}$.
$f_1^{\mathrm{odd}}$ represents an induced pairing correlation belonging to the odd-frequency symmetry class
and is finite for $P_{\mathrm{O}} \neq 0$~\cite{kim:prb2021,kim:jpsj2023}.
The structure of $f_1^{\mathrm{q}}$ is modified by $f_1^{\mathrm{odd}}$
because the two correlation functions are related to each other through 
the Gor'kov equation in Eq.~\eqref{eq:gorkov}.
Therefore, the pseudospin-quintet components linked to the pair potential
are indirectly modified by the odd-frequency component.
The singlet component $f_1^{\mathrm{s}}$ and odd-frequency component $f_1^{\mathrm{odd}}$ do not form 
any pair potentials because the attractive interactions for the corresponding pairing channels are 
absent at the starting Hamiltonian.

In the absence of the odd-frequency pairing correlations (i.e., $P_{\mathrm{O}} = 0$),
 the coefficients in the free-energy functional are calculated to be
\begin{align}
    \label{eq:fq_w/o_odd}
    f_1^{\mathrm{q}} (\bm{k},i\omega_n) &=
    -\vec{\epsilon}_{\bm{k}}^{\,2} \, \vec{\eta}_{\bm{k}} \cdot \vec{\gamma} \, U_T , \\
    a &=
    \label{eq:a_w/o_odd}
    \frac{1}{\tilde{g}} + T\sum_{\omega_n} \frac{1}{N}\sum_{\bm{k}}
    \frac{-2}{Z_0} ( \omega_n^2 + \xi_{\bm{k}}^2 + \vec{\epsilon}_{\bm{k}}^{\,2} ) , \\
    b &=
    \label{eq:b_w/o_odd}
    T\sum_{\omega_n} \frac{1}{N}\sum_{\bm{k}}
    \frac{1}{Z_0^2}
    \left\{
        ( \omega_n^2 + \xi_{\bm{k}}^2 + \vec{\epsilon}_{\bm{k}}^{\,2} )^2
        + 4 \xi_{\bm{k}}^2 \vec{\epsilon}_{\bm{k}}^{\,2}
    \right\}
   \left\{
        2 - |\vec{\eta}_{\bm{k}} \cdot \vec{\eta}_{\bm{k}}|^2
    \right\} .
%
\end{align}
    $\vec{\epsilon}_{\bm{k}}^{\,2}$ in the last term of the numerator of Eq.~\eqref{eq:a_w/o_odd} originates from $f_1^{\mathrm{q}}$
    and amplifies the integrand.
The gap equation corresponding to $a=0$ has the same expression as
that in the BCS theory~\cite{kim:jpsj2023}.
In addition, the coefficient $b$ is always positive.
As a result, the transition to the superconducting state is a second-order 
and the superconducting state is stable for $T<T_c$.
Therefore, the equation $P_{\mathrm{O}}=0$ characterizes the perturbations that preserve the thermal properties 
of the superconducting states.
Namely, the thermal properties of superconducting states in the absence of odd-frequency Cooper pairs 
are identical to those in the BCS state.
We note that the thermal properties of a pseudospin-singlet superconducting state are also identical to those in 
the BCS state as shown in Appendix.~\ref{sec:singlet_j=3/2}.

In the presence of odd-frequency pairing correlation (i.e., $P_{\mathrm{O}} \neq 0$),
it is not easy to obtain analytical expression of the GL coefficients
without further simplifications.
To proceed discussions, we restrict ourselves to consider $E_g$ pairing order
$\vec{\eta}_{\bm{k}} = \left( 0, 0, 0, \eta_{\bm{k},4}, \eta_{\bm{k},5} \right)$
because there are only two components in the pair potential.
This simplification enables us to get the following analytical expressions:
\begin{align}
    \label{eq:fq_w/_odd}
    f_1^{\mathrm{q}} (\bm{k},i\omega_n) &=
    \vec{\epsilon}_{\bm{k}}^{\,2} \, \vec{\eta}_{\bm{k}} \cdot \vec{\gamma} \, U_T
    - 2 \vec{\epsilon}_{\bm{k}} \cdot \vec{\eta}_{\bm{k}} \, \vec{\epsilon}_{\bm{k}} \cdot \vec{\gamma} \, U_T , \\
    \label{eq:a_w/_odd}
    a &=
    \frac{1}{\tilde{g}} +
    T\sum_{\omega_n} \frac{1}{N}\sum_{\bm{k}}
    \frac{-2}{Z_0}
    ( \omega_n^2 + \xi_{\bm{k}}^2 + \vec{\epsilon}_{\bm{k}}^{\,2} - 2 A_0 ), 
    \quad A_0 = \vec{\epsilon}_{\bm{k}}^{\,2} - |\vec{\epsilon}_{\bm{k}} \cdot \vec{\eta}_{\bm{k}}|^2,\\
    \label{eq:b_w/_odd}
    b &=
    T\sum_{\omega_n} \frac{1}{N}\sum_{\bm{k}}
    \frac{1}{Z_0^2}
    \bigg[
        ( \omega_n^2 + \xi_{\bm{k}}^2 - \vec{\epsilon}_{\bm{k}}^{\,2} )^2
        -4\omega_n^2 \vec{\epsilon}_{\bm{k}}^{\,2} \nonumber \\
    &\phantom{{}={} \bigg[} -
    \left\{
        ( \omega_n^2 + \xi_{\bm{k}}^2 - \vec{\epsilon}_{\bm{k}}^{\,2} )^2
        -4\omega_n^2
        \left(
            \epsilon_{\bm{k},1}^{\,2} + \epsilon_{\bm{k},2}^{\,2} + \epsilon_{\bm{k},3}^{\,2}
            -\epsilon_{\bm{k},4}^{\,2} - \epsilon_{\bm{k},5}^{\,2}
        \right)
    \right\}
    \eta_{\bm{k},4}^2 \left( \eta_{\bm{k},5} - \eta_{\bm{k},5}^{\ast} \right)^2
    \nonumber \\
    &\phantom{{}={} \bigg[} +
    8|\vec{\epsilon}_{\bm{k}} \cdot \vec{\eta}_{\bm{k}}|^2
    \left(
        \omega_n^2 + \xi_{\bm{k}}^2 - \vec{\epsilon}_{\bm{k}}^{\,2} +
        |\vec{\epsilon}_{\bm{k}} \cdot \vec{\eta}_{\bm{k}}|^2
    \right)
    \bigg],
\end{align}
where we chose the common phase factor of the pair potential so that $\Delta$ and $\eta_{\bm{k},4}$ are real
but $\eta_{\bm{k},5}$ is complex in general.
\end{widetext}
We found that the expression of $f_1^{\mathrm{q}}$ and $a$ in Eqs.~\eqref{eq:fq_w/_odd} and \eqref{eq:a_w/_odd} is also valid for the general case:
$\vec{\eta}_{\bm{k}} = \left( \eta_{\bm{k},1}, \eta_{\bm{k},2}, \eta_{\bm{k},3}, \eta_{\bm{k},4}, \eta_{\bm{k},5} \right)$.
Comparing with Eq.~\eqref{eq:fq_w/o_odd}, the sign of the first term in Eq.~\eqref{eq:fq_w/_odd} is 
reversed due to $f_1^{\mathrm{odd}}$
and a correction appears at the second term.
Comparing with Eq.~\eqref{eq:a_w/o_odd}, an additional term $-2A_0$ 
appears in Eq.~\eqref{eq:a_w/_odd} to compensate for the presence of odd-frequency pairs.
Since $A_0 \geqq 0$,
the odd-frequency component $f_1^{\mathrm{odd}}$ suppresses the amplitudes of the pseudospin-quintet 
components, which leads to the suppression of $T_c$~\cite{asano:prb2015}.
Similar arguments have also been presented in other papers~\cite{ramires:prb2016,ramires:prb2018,triola:annphys2020}.
The paramagnetic property of uniform odd-frequency Cooper pairs are summarized in Appendix.~\ref{sec:paraodd}.
Even under the simplifications in the pair potential 
$\vec{\eta}_{\bm{k}} = \left( 0, 0, 0, \eta_{\bm{k},4}, \eta_{\bm{k},5} \right)$, 
$P_{\mathrm{O}} \neq 0$ makes the expression of the coefficient $b$ lengthy and complicated
as shown in Eq.~\eqref{eq:b_w/_odd}.
To obtain the physical insights from the analytical expression of $b$, we consider specific examples such as
$\left( \eta_{\bm{k},4},\eta_{\bm{k},5} \right) = (1,0)$, $(0,1)$, $(1,1)/\sqrt{2}$, and $(1,i)/\sqrt{2}$.
The first, second, and fourth ones are predicted to be stable states within the phenomenological GL theory~\cite{sigrist:rmp1991,brydon:prl2016,brydon:prb2018}.
The commutator in Eq.~\eqref{eq:Po_j32} is calculated for $(1,0)$ state 
\begin{align}
    P_{\mathrm{O}}^{(1,0)} &= 2 \gamma^4 \sum_{\mathrm{i} \neq 4} \epsilon_{\bm{k}, \mathrm{i}} \gamma^{\mathrm{i}} .
\end{align}
The results for $(0,1)$, $(1,1)/\sqrt{2}$, and $(1,i)/\sqrt{2}$ states
are given in Appendix~\ref{sec:triplet-septet}.
The resulting correlation functions in Eq.~\eqref{eq:Po_j32} have a common matrix structure
\begin{align}
f_1^{\mathrm{odd}}(\bm{k}, i\omega_n) \propto i \omega_n \, \sum_{\mathrm{i}=4,5} \sum_{\mathrm{j}\neq\mathrm{i}}
a_{\mathrm{i,j}}(\bm{k}) \, \gamma^{\mathrm{i}}\, \gamma^{\mathrm{j}}\, U_T,
\label{eq:induced_odd}
\end{align}
where $a_{\mathrm{i,j}}$ is an even-parity function. 
Since $[f_1^{\mathrm{odd}}(\bm{k}, i\omega_n)]^\mathrm{T} =f_1^{\mathrm{odd}}(\bm{k}, i\omega_n)$, the induced 
odd-frequency pairing correlations consist of pseudospin-triplet states and pseudospin-septet states.

The expression of the coefficient $b$ 
\begin{align}
    b^{\mathrm{TRS}} &=
    T\sum_{\omega_n} \frac{1}{N}\sum_{\bm{k}}
    \frac{1}{Z_0^2}
    \bigg[
        ( \omega_n^2 + \xi_{\bm{k}}^2 - \vec{\epsilon}_{\bm{k}}^{\,2} )^2 \nonumber\\
  &    + 4|\vec{\epsilon}_{\bm{k}} \cdot \vec{\eta}_{\bm{k}}|^2
        \left(
            \omega_n^2 + 2\xi_{\bm{k}}^2 - 2 A_0
        \right)
        -4 \omega_n^2 A_0
    \bigg]  \label{eq:btrs_gen} , 
\end{align}
is common in all the time-reversal invariant states
$\left( \eta_{\bm{k},4},\eta_{\bm{k},5} \right) = (1,0)$, $(0,1)$, and $(1,1)/\sqrt{2}$.
The results of $b$ for a time-reversal breaking state $(1,i)/\sqrt{2}$ are given in
Appendix~\ref{sec:triplet-septet}.
The last terms proportional to $\omega_n^2$ in Eqs.~\eqref{eq:btrs_gen} and \eqref{eq:btrsb_gen}
originate from $f_1^{\mathrm{odd}}$ and are always less than or equal to zero.
Thus, the last term decreases the coefficient $b$ down to zero when the amplitude of $f_1^{\mathrm{odd}}$ is 
sufficiently large.
The relation $P_{\mathrm{O}} \neq 0$ in Eq.~\eqref{eq:Po_j32} 
characterizes perturbations that change the thermal properties of the superconducting states.
Therefore, the two types of perturbations mentioned in the introduction 
are distinguished by whether or not they induce odd-frequency Cooper pairs.

The sixth and higher order terms in Eq.~\eqref{eq:gl_func} are also modified by the odd-frequency 
correlation functions.
However, it is not easy to separate the contributions of the odd-frequency correlations 
from those of the even-frequency correlations because of their cross terms. 
An example of the analytical expression of the sixth-order coefficient $c$
is given in Appendix.~\ref{sec:sixth_coef}.

\section{Discontinuous transition}
\label{sec:discontinuous}
\begin{figure}[tbp]
    \includegraphics[width=8.5cm]{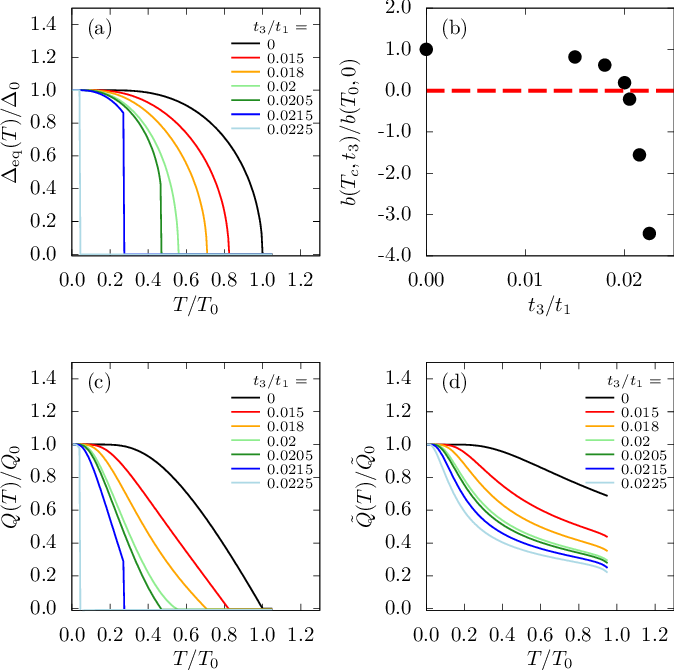}
    \caption{
        The self-consistent solution of the pair potential $\Delta_{\mathrm{eq}}(T)$ in a $j=3/2$ superconductor
        is plotted as a function of temperature
        for several strengths of spin-orbit interaction $t_3$ in (a), where
        $T_0$ is the transition temperature at $t_3=0$ and $\Delta_0$ is the amplitude of the pair potential at $T=0$ and $t_3=0$.
        The coefficient $b$ at $T=T_c$ is plotted as a function of $t_3$ in (b),
        where $T_c$ is obtained from the results in (a).
        The temperature dependence of
        the superfluid density $Q$ and $\tilde{Q}$ is shown in (c) and (d), respectively.
        $Q_0$ ($\tilde{Q}_0$) in (c) ((d)) represents $Q$ ($\tilde{Q}$) at $T=0$ and $t_3=0$.
    }
    \label{fig:delta_q_b_j32}
\end{figure}

In this section, we demonstrate that the transition to the superconducting phase becomes discontinuous
when the amplitude of the odd-frequency pairing components are sufficiently large.
We choose $t_2=0$ in the normal state Hamiltonian in Eq.~\eqref{eq:hn}.
This simplification enables us to solve the Gor'kov equation in Eq.~\eqref{eq:gorkov} 
analytically up to the infinite order of $\Delta$.
In what follows, we consider a pair potential 
$\left( \eta_{\bm{k},4},\eta_{\bm{k},5} \right) = (1,0)$ preserving time-reversal symmetry. 
The following discussions are valid also for $\left( \eta_{\bm{k},4},\eta_{\bm{k},5} \right) =(0,1)$, 
$(1,1)/ \sqrt{2}$, and $(1,i)/\sqrt{2}$.
We found that the thermal properties are the same among these states at $t_2=0$.
The existence of odd-frequency Cooper pairs is a common feature among these states.
The anomalous Green's function for $\left( \eta_{\bm{k},4},\eta_{\bm{k},5} \right) = (1,0)$  results in
\begin{align}
    &\mathcal{F}(\bm{k},i\omega_n) =
    -\frac{\Delta}{Z}
    \left[
        W -2i\omega_n \vec{\epsilon}_{\bm{k}} \cdot \vec{\gamma}
    \right]
    \vec{\eta}_{\bm{k}} \cdot \vec{\gamma} U_{T} , \label{eq:f-func_main}\\
    &Z = W^2 + 4\omega_n^2 \vec{\epsilon}_{\bm{k}}^{\,2}, \quad 
    W=     \omega_n^2 + \xi_{\bm{k}}^2 - \vec{\epsilon}_{\bm{k}}^{\,2}  + \Delta^2.\label{eq:zw_main}
\end{align}
The second term in Eq.~\eqref{eq:f-func_main} is the pairing correlation belonging to the
odd-frequency 
symmetry class, which is induced by the spin-orbit interaction $\vec{\epsilon}_{\bm{k}}$.
The coefficient of the fourth-order term is calculated to be
\begin{align}
	b(T) &= T\sum_{\omega_n} \frac{1}{N}\sum_{\bm{k}}
    \frac{1}{Z_0^2} 
    \left[
       W_0^2        -4 \omega_n^2 \vec{\epsilon}_{\bm{k}}^{\,2}
    \right] ,\label{eq:btrs}
\end{align}
with $Z_0 = \left. Z\right|_{\Delta=0}$ and 
$W_0 = \left. W\right|_{\Delta=0}$.
The last term in Eq.~\eqref{eq:btrs} is derived from the odd-frequency pairing correlation 
functions and contributes negatively to the coefficient $b$ as already mentioned in 
Eq.~\eqref{eq:btrs_gen}.
The amplitude of the pair potential $\Delta$ is determined self-consistently from the
thermodynamic potential in the superconducting state
\begin{align}
    \Omega_{\mathrm{S}}(\Delta)
    &=
    \frac{\Delta^2}{\tilde{g}} - \frac{2T}{N} \sum_{\bm{k},\lambda=\mathrm{S}\pm}
        \log \left[ 2\cosh \left( \frac{E_{\lambda}(\bm{k})}{2T} \right) \right] ,
%
\end{align}
where 
$E_{\mathrm{S} \pm}(\bm{k}) = \sqrt{\xi_{\bm{k}}^2 + \Delta^2} \pm |\vec{\epsilon}_{\bm{k}}|$ 
and irrelevant constants are neglected.
The pair potential is determined
by minimizing $\Omega_{\mathrm{S}}(\Delta)$ with respect to $\Delta$.
Thus, the solution in the equilibrium state ($\Delta_{\mathrm{eq}}$) always satisfies
\begin{align}
    \Omega_{\mathrm{SN}}(\Delta_{\mathrm{eq}}) &=
    \min_{\Delta} \left\{ \Omega_{\mathrm{SN}} (\Delta) \, | \, \Delta \in \mathbb{R} \right\}
    \leqq 0 ,
\end{align}
with $\Omega_{\mathrm{SN}}(\Delta) \equiv \Omega_{\mathrm{S}}(\Delta) - \Omega_{\mathrm{S}}(0)$.
The solution of $\Delta_{\mathrm{eq}}$ is plotted as a function of temperature
in Fig.~\ref{fig:delta_q_b_j32}(a) for several choices of spin-orbit interaction $t_3$.
Hereafter, the transition temperature at $t_3=0$ is denoted by $T_0$, and the pair potential 
at $T=0$ and $t_3=0$ is denoted by $\Delta_0$.
In the numerical simulation, we chose
$\mu = t_1$ and
$\tilde{g} = 2.463 t_1$ so that $T_0 = 0.05 t_1$.
We obtained $\Delta_0 = 0.0882 t_1 \approx 1.76 T_0$,
which corresponds to BCS universal relation~\cite{tinkham}.
The transition temperature decreases monotonically with increasing $t_3$.
Although $\Delta_{\mathrm{eq}}$ is insensitive to $t_3$ at very low temperature $T\ll T_0$, 
it abruptly vanishes for $t_3 \gtrsim 0.023 t_1$.
 Uniform superconducting states are no longer stable under strong spin-orbit couplings.
Furthermore, $\Delta_{\mathrm{eq}}$ shows the discontinuous behavior at $T_c$ for 
$t_3 \gtrsim 0.0205 t_1$.
In Fig.~\ref{fig:delta_q_b_j32}(b), the coefficient $b(T_c)$
is plotted as a function of $t_3$.
We obtained $b(T_0) = 1.237 t_1^{-3}$ at $t_3 = 0$.
As predicted in Eq.~\eqref{eq:btrs}, the odd-frequency pairing correlations decrease 
the coefficient $b(T_c)$.
As a result, the transition becomes discontinuous for $b(T_c)<0$ as shown in 
Fig.~\ref{fig:delta_q_b_j32}(a) and (b).
Thus, odd-frequency Cooper pairs are responsible for the 
discontinuous transition to the superconducting states.

\section{Superfluid density}
\label{sec:pairdens}
To understand why odd-frequency Cooper pairs cause the discontinuous transition, 
we discuss the relationship between the coefficient $b$ and the response 
function to an electromagnetic field,
\begin{align}
    j_{x} (\bm{q},\omega) = - K_{xx} (\bm{q},\omega) A_x (\bm{q},\omega) ,
\end{align}
where $j_\nu$ is the electric current and $A_\nu (\bm{q},\omega)$ is the Fourier component of a vector potential.
The derivation of the response kernel $K_{\nu\nu}$ is presented in 
Appendix~\ref{sec:em_response}.
The response kernel to a static transverse gauge potential is called
 Meissner kernel or \textsl{superfluid density}
\begin{align}
    Q= \frac{K_{xx}( \bm{q} \rightarrow 0, \omega=0 )}{2e^2 t_1} ,
\end{align}
%
%
where $e$ is the charge of an electron and $Q$ has no dimension.
%
%
In Fig.~\ref{fig:delta_q_b_j32}(c), the superfluid density $Q$ is plotted as a function of temperature
for several choices of $t_3$, where $Q_0= 0.0664$ is the superfluid density
at $T = 0$ and $t_3 = 0$ in our numerical simulation.
At $T \approx 0$, the superfluid density is almost independent of $t_3$ for $t_3 \lesssim 0.0225 t_1$.
However, the superfluid density decreases drastically at finite temperatures.
To understand such characteristic features,
we analyze the contribution of the anomalous Green's function in Eq.~\eqref{eq:f-func_main}
to the superfluid density,
\begin{align}
    Q^{\mathcal{F}} &=
    T\sum_{\omega_n} \frac{1}{N}\sum_{\bm{k}}
    2t_1 \sin^2 k_x \frac{4\Delta^2}{Z^2}\left[
    W^2 - 4 \omega_n^2 \, \vec{\epsilon}_{\bm{k}}^{\,2}
    \right] . \label{eq:qf}
\end{align}
The second term is derived from the odd-frequency pairing correlations
and reduces the superfluid density.
The dependence of $Q$ on temperature in Fig.~\ref{fig:delta_q_b_j32}(c)
is dominated mainly by that of $\Delta_{\mathrm{eq}}^2(T)$
because $Q$ is proportional to $\Delta_{\mathrm{eq}}^2(T)$ as shown in Eq.~\eqref{eq:qf}.
Thus, it is not easy to extract the effects of odd-frequency pairs on the superfluid density.
To highlight a role of odd-frequency pairs in the discontinuous transition, we calculate
\begin{align}
    \tilde{Q}(T, t_3) &= \frac{Q(T, t_3, \Delta_{\mathrm{BCS}}(T))}{\Delta_{\mathrm{BCS}}^2(T)} ,
\end{align}
for $T < T_0$.
Here we first replace $\Delta_{\mathrm{eq}}(T, t_3)$ by
\begin{align}
    \Delta_{\mathrm{BCS}}(T) = \Delta_{\mathrm{eq}}(T, t_3=0) ,
\end{align}
and divide the results by $\Delta_{\mathrm{BCS}}^2(T)$ to relax the influence of $\Delta_{\mathrm{BCS}}(T)$.
In Fig.~\ref{fig:delta_q_b_j32}(d), $\tilde{Q}$ is plotted for several choices of $t_3$.
The vertical axis is normalized to $\tilde{Q}_{0}=\tilde{Q}(T=0, t_3=0)$.
The black line for $t_3=0$ almost corresponds to the results of BCS theory
\begin{align}
    \frac{\tilde{Q}(T, t_3=0)}{\tilde{Q}_0} \approx
    \Delta_0^2 \, \pi T \sum_{\omega_n} \frac{1}{( \omega_n^2 + \Delta_{\mathrm{BCS}}^2(T))^{3/2}} ,
\end{align}
and decreases with increasing temperature almost linearly for $T \gtrsim 0.3 T_0$. 
$\tilde{Q}$ at $T=0$ remains unchanged even in the presence of the spin-orbit interaction,
whereas it at finite temperatures decreases with increasing $t_3$.
The suppression from the black line is remarkable for $0.2 \lesssim T/T_0 \lesssim 0.5$.
As a result, the curves for $t_3/t_1 = 0.015 - 0.0225$ are convex downward.
The drastic suppression of the superfluid density in such finite temperatures
is responsible for the suppression of $T_c$
and the discontinuous transition finding at $t_3/t_1 \gtrsim 0.0205$.

When we compare Eq.~\eqref{eq:btrs} with Eq.~\eqref{eq:qf}, 
the odd-frequency pairs decrease the coefficient $b$ and the superfluid density $Q^{\mathcal{F}}$
in the same manner.
The two values are related to each other by the relation
\begin{align}
b \approx A\,  \left. Q^{\mathcal{F}} / \Delta^2 \right|_{\Delta\to 0}, \label{eq:bq-relation}
\end{align}
with $A>0$ being a constant.
$Q^{\mathcal{F}}$ in Eq.~\eqref{eq:bq-relation} would be replaced by $Q$ (total superfluid density)
if we can perform the momentum integration analytically.
The relationship between $b$ and $Q$ in Eq.~\eqref{eq:bq-relation} at the discontinuous transition 
is a central finiding in this paper. We will revisit the relation in other superconducting 
states in Sec.~\ref{sec:zeeman}.
As shown in Eq.~\eqref{eq:f-func_main}, the odd-frequency pairing correlation function
is proportional to the Matsubara frequency, which is a common property of odd-frequency pairs 
in uniform superconductors~\cite{asano:prb2015,triola:annphys2020}.
As a result, the instability due to odd-frequency pairs is considerable 
at finite temperatures~\cite{sasaki:prb2020}.
We conclude that the discontinuous transition to the superconducting phase occurs because
odd-frequency Cooper pairs reduce the superfluid density at finite temperatures.
%

\section{Universality of phenomenon}
\label{sec:zeeman}

\subsection{Other $j=3/2$ states}

The discontinuous transition due to odd-frequency Cooper pairs and 
the relationship in Eq.~\eqref{eq:bq-relation} are
confirmed in other superconducting states.
Indeed, the expression in Eqs.~\eqref{eq:btrs} and \eqref{eq:qf} can be applied also to
other $E_g$ states such as
$\left( \eta_{\bm{k},4},\eta_{\bm{k},5} \right) =(0,1)$ and $(1,1)/\sqrt{2}$
at $t_2=0$.

The discontinuous transition in $j=3/2$ superconductors has also been reported 
in $T_{2g}$ pairing states at $T \geqq 0$~\cite{menke:prb2019}.
The authors of Ref.~\cite{menke:prb2019} concluded that the interband pairs
are responsible for the discontinuous transition.
Here, \textsl{band} means the diagonalized normal state band (i.e., Bloch band).
    The pair potentials for such $T_{2g}$ states can be described as
    \begin{align}
        \vec{\eta}_{\bm{k}} &= (0,\eta_{\bm{k},2},\eta_{\bm{k},3},0,0).
\end{align}
	The odd-frequency Cooper pairs exist also in $T_{2g}$ states. 
Since the commutator in Eq.~\eqref{eq:Po_j32} for $T_{2g}$ states  
is calculated to be
\begin{align}
		P_{\mathrm{O}}^{T_{2g}} &= 2
        \left(
            \eta_{\bm{k},2} \gamma^2
            \sum_{\mathrm{i} \neq 2} \epsilon_{\bm{k}, \mathrm{i}} \gamma^{\mathrm{i}}
            +
            \eta_{\bm{k},3} \gamma^3
            \sum_{\mathrm{i} \neq 3} \epsilon_{\bm{k}, \mathrm{i}} \gamma^{\mathrm{i}}
        \right),
    \end{align}
the odd-frequency pairing correlation function has a mathematical structure of
\begin{align}
f_1^{\mathrm{odd}}(\bm{k}, i\omega_n) \propto i \omega_n \, \sum_{\mathrm{i}=2,3} \sum_{\mathrm{j}\neq\mathrm{i}}
a_{\mathrm{i,j}}(\bm{k}) \, \gamma^{\mathrm{i}}\, \gamma^{\mathrm{j}}\, U_T.
\label{eq:t2g_odd}
\end{align}
The correlation function has essentially the same structures as that in Eq.~\eqref{eq:induced_odd}.
    Such an analysis suggests that the discontinuous transition due to odd-frequency Cooper pairs 
	also occurs in the $T_{2g}$ states.
	Unfortunately, however, we cannot conclude clearly because we cannot
 derive the relation in Eq.~\eqref{eq:bq-relation} analytically. 

   In Ref.~\cite{menke:prb2019}, the authors pointed out an important role of interband 
   Cooper pairs in the discontinuous transition. Thus it seems meaningful to compare
   the two pairing states and sort out their relationship.
   The interband pairing and odd-frequency pairing
   are not equivalent concept to each other. To make this point clear, we consider a two-band 
   superconductor described by
    \begin{align}
        H_{\mathrm{BdG}}(\bm{k}) &=
        \left[
            \begin{array}{cc}
                E_{\mathrm{N}}(\bm{k}) & \Delta(\bm{k}) \\
                \Delta(\bm{k}) & -E_{\mathrm{N}}(\bm{k}) 
            \end{array} 
        \right] , \\
		 E_{\mathrm{N}} &=
		         \left[
            \begin{array}{cc}
               \epsilon_1 & 0 \\
                0 & \epsilon_2
            \end{array} 
        \right] , \quad 
		 \Delta=
		         \left[
            \begin{array}{cc}
               \Delta_1 & 0 \\
                0 & \Delta_2
            \end{array} 
        \right].		
    \end{align}
where the normal state Hamiltonian is diagonalized by a unitary transformation. 
We assume that the pair potential has only diagonal elements in the Bloch band picture. 
The spin symmetry and the momentum-parity of the pair potentials $\Delta_j$ can be either
spin-singlet even-parity or spin-triplet odd-parity. 
We assume $E_{\mathrm{N}}(-\bm{k})=E_{\mathrm{N}}(\bm{k})$ for simplicity.
The BdG Hamiltonian can be block-diagonalized for each band.
Therefore, the interband Cooper pairs are absent in such a superconducting state. 
The odd-frequency pairing correlation in this case 
$
f^{\mathrm{odd}} \propto i \omega_n [ E_{\mathrm{N}} \,\Delta - \Delta \,E_{\mathrm{N}}]$
is also absent because the commutator is zero. 
Next, we introduce the interband superconducting order parameter  
\begin{align}
		 \Delta=
		         \left[
            \begin{array}{cc}
               \Delta_1 & \Delta_{12} \\
                \Delta_{12} & \Delta_2
            \end{array} 
        \right]. \label{eq:interband2}		
    \end{align}
Here we assume that $\Delta_{12}$ appears as a results of the unitary transformation 
that diagonalizes the normal state Hamiltonian.
Three pairing correlations contribute to the pair potential in Eq.~\eqref{eq:interband2}. 
In addition to the two intraband pairing correlations, the interband pairing correlation forms
 the pair potentials.
The pair potential in Eq.~\eqref{eq:interband2} satisfies $\Delta^{\mathrm{T}}=\Delta$, which 
means the pair potentials are symmetric under interchanging the band indices.
When the two bands are identical to each other $\epsilon_1=\epsilon_2$, 
 $E_{\mathrm{N}}$ is proportional to the identity matrix and commutes with $\Delta$.
As a result, the odd-frequency pairing correlation is absent. 
For $\epsilon_1\neq \epsilon_2$, the asymmetry between the two bands generates 
the odd-frequency pairs as a subdominant pairing correlation
\begin{align}
f_{\mathrm{odd}} \propto i \omega_n 
		         \left[
            \begin{array}{cc}
               0 & \Delta_{12}(\epsilon_1 - \epsilon_2) \\
                -\Delta_{12}(\epsilon_1 - \epsilon_2) & 0
            \end{array} 
        \right].	
\end{align}
In contrast to the interband pairing correlation forming $\Delta_{12}$ in Eq.~\eqref{eq:interband2}, 
the induced interband pairing correlation is antisymmetric under interchanging
band indices, (i.e., $f_{\mathrm{odd}}^{\mathrm{T}} =- f_{\mathrm{odd}}$).	 
Cooper pairs linked to $\Delta_{12}$ belong to even-frequency even-band-parity symmetry class,
whereas induced Cooper pairs belong to odd-frequency odd-band-parity symmetry class.
Such symmetry conversion occurs because the band asymmetry preserves both spin configurations 
and momentum parity of a Cooper pair. 
Thus, the interband Cooper pairing and the odd-frequency Cooper pairing are different concepts from each other.
In Ref.~\cite{menke:prb2019}, the authors may not distinguish two interband Cooper pairs.
However, in this paper, we clearly distinguish the two interband pairs because 
they contribute to the superfluid density in opposite ways to each other.

The instability at $T=0$ for both $T_{2g}$ and $E_g$ pairing states has been
also discussed in Ref.~\cite{bhattacharya:prb2023}.
The authors of Ref.~\cite{bhattacharya:prb2023} compared the free-energy 
among multiple superconducting states with changing the amplitude of the attractive 
interaction between the two electrons.
They found that the transition to 
another superconducting phase becomes first-order. 
Odd-frequency pairing correlations are also present in these superconducting states.
Therefore, the transition from the normal state to one of these superconducting states  
would be discontinuous when the amplitude of the odd-frequency pairs are sufficiently large.
%

\begin{table*}[t]
\caption{
Three theoretical models which describe the discontinuous transition to the superconducting phase.
The structures of the pair potential are given in the second row.
The third row shows the perturbations that induce odd-frequency pairing correlations.
The transition to the superconducting state becomes discontinuous because 
odd-frequency Cooper pairs decrease the superfluid density to zero. 
The fourth order coefficient of the GL free-energy $b$ and 
the superfluid density $Q$ are proportional to each other as shown in Eq.~\eqref{eq:bq-relation}.
 }
\begin{ruledtabular}
\renewcommand{\arraystretch}{1.3}
\begin{tabular}{cccc}
\null &
 j=3/2 SC & conventional SC & two-band SC\\
\colrule
pair potential & $\Delta \vec{\eta}_{\bm{k}} \cdot \vec{\gamma} \, U_T $ & $\Delta \, i \hat{\sigma}_2$ 
& $\Delta \, i\hat{\rho}_2$\\
\colrule
perturbation &
\begin{tabular}{c}
    spin-orbit interaction \\ $\vec{\epsilon}_{\bm{k}} \cdot \vec{\gamma} $
\end{tabular} &
\begin{tabular}{c}
    Zeeman field \\ $\mu_{\mathrm{B}} \bm{B} \cdot \hat{\bm{\sigma}}$
    \footnote{$\hat{\sigma}_\mathrm{j}$ for $\mathrm{j}=1-3$ are Pauli matrices in spin space.}
\end{tabular} &
\begin{tabular}{c}
    band hybridization \& asymmetry \\ $\epsilon \, \hat{\rho}_3 + V \,\hat{\rho}_1$
    \footnote{$\hat{\rho}_\mathrm{j}$ for $\mathrm{j}=1-3$ are Pauli matrices in band space.}
\end{tabular} \\
\colrule
$b \approx A\, Q^{(\mathcal{F})}/\Delta^2|_{\Delta \to 0}$ & Eqs.~\eqref{eq:btrs} and \eqref{eq:qf} & 
Eqs.~\eqref{eq:b_zeeman_text} and \eqref{eq:q_zeeman_main_text} & Eq.~\eqref{eq:b_q_interband}\\
\end{tabular}
\end{ruledtabular}
\label{table1}
\end{table*}

\begin{figure}[tbp]
    \includegraphics[width=8.5cm]{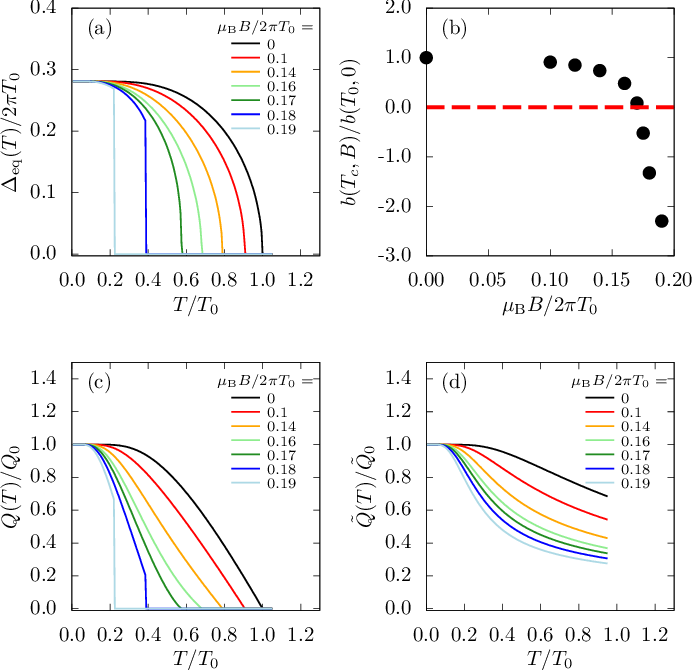}
    \caption{
        The self-consistent solution of the pair potential $\Delta_{\mathrm{eq}}(T)$ in a spin-singlet superconductor under a Zeeman field
        is plotted as a function of temperature
        for several strengths of Zeeman interaction $B$ in (a), where
        $T_0$ is the transition temperature at $B=0$ and $\Delta_0$ is the amplitude of the pair potential at $T=0$ and $B=0$.
        The coefficient $b$ at $T=T_c$ is plotted as a function of $B$ in (b),
        where $T_c$ is obtained from the results in (a).
        The temperature dependence of
        the superfluid density $Q$ and $\tilde{Q}$ is shown in (c) and (d), respectively.
        $Q_0$ ($\tilde{Q}_0$) in (c) ((d)) represents $Q$ ($\tilde{Q}$) at $T=0$ and $B=0$.
    }
    \label{fig:delta_q_b_zeeman}
\end{figure}

\subsection{Other $s=1/2$ states}

In addition to $j=3/2$ superconductors,
the discontinuous transition has been found in other superconducting states of $s=1/2$ electrons.
The transition in a two-band superconductor
with interband pairing order becomes one of the examples
when the band hybridization $V$ is sufficiently large~\cite{silva:physletta2014}.
We note that \textsl{bands} in Ref.~\cite{silva:physletta2014} indicates atomic orbitals rather than the Bloch bands.
In this model, the band-hybridization corresponds to the perturbation which generates odd-frequency Cooper pairs. 
The paramagnetic property of the odd-frequency Cooper pairs explains well the mechanism of the discontinuous transition.
The analysis for such an interband superconductor in Ref.~\cite{silva:physletta2014}
is presented in Appendix~\ref{sec:interband}.
We also discuss the effects of band asymmetry $\epsilon$ on the discontinuous transition.
The fourth-order coefficient of the GL free-energy $b$
and the superfluid density $Q$ share the same expression as shown in Eq.~\eqref{eq:b_q_interband}.
As a result, we confirm the relation in Eq.~\eqref{eq:bq-relation} also in a two-band superconductor.

Finally, we emphasize the relevance of the conclusions in this paper to an important open issue. 
The transition to the uniform spin-singlet $s$-wave superconducting state is known to be
discontinuous under a Zeeman field $B$~\cite{sarma:jpcs1963,maki_tsuneto:ptp1964,chandrasekhar:apl1962,clogston:prl1962}.
The calculated results for the coefficient $b$ and the superfluid density $Q$ are given by
    \begin{align}
        &b = \frac{N_0}{4} Y(A_0, C_0) , \label{eq:b_zeeman_text}\\
        &Q = 2n \Delta^2 Y(A, C) , \label{eq:q_zeeman_main_text}\\
        &Y(A,C) =\sqrt{2} \pi T \sum_{\omega_n}
        \frac{A^3+\sqrt{C}(A^2-2\omega_n^2 \mu_{\mathrm{B}}^2 B^2)}
        {[ C (A+\sqrt{C})]^{3/2}} , \label{eq:yac} \\
        &A=\Delta^2 + \omega_n^2-\mu_{\mathrm{B}}^2 B^2, \;  C=A^2 + 4 \omega_n^2\, \mu_{\mathrm{B}}^2 B^2 , \label{eq:Xp}
    \end{align}
with $A_0=A|_{\Delta=0}$ and $C_0=C|_{\Delta=0}$, where 
$N_0$ is the density of states at the Fermi level per spin in the normal state, $n$ is the electron density per spin, 
and $\mu_{\mathrm{B}}$ is Bohr's magneton.
The derivations are given in Appendix.~\ref{sec:singlet}.
We also calculate
\begin{align}
    \tilde{Q}(T, B) &= \frac{Q(T, B, \Delta_{\mathrm{BCS}}(T))}{\Delta_{\mathrm{BCS}}^2(T)} \nonumber \\
    &= 2n Y(A, C)|_{\Delta = \Delta_{\mathrm{BCS}}(T)} ,
\end{align}
where $\Delta_{\mathrm{BCS}}(T) = \Delta_{\mathrm{eq}}(T, B=0)$
represents the pair potential of a BCS superconductor.
The last term in Eq.~\eqref{eq:yac} is derived from the odd-frequency pairing correlation,
which is generated by a Zeeman field.
The coefficient $b$ and the superfluid density $Q$ satisfy the relation in Eq.~\eqref{eq:bq-relation}.
The self-consistent pair potential $\Delta_{\mathrm{eq}}$,
the coefficient $b$ at the transition temperature,
the superfluid density $Q$, and $\tilde{Q}$
in the spin-singlet superconductor are plotted
in Fig.~\ref{fig:delta_q_b_zeeman}(a), (b), (c), and (d), respectively.
We denote the transition temperature at $B=0$ by $T_0$
and the pair potential at $T=0$ and $B=0$ by $\Delta_0 \approx 1.76 T_0$~\cite{tinkham}.
The coefficient $b$ at $T=T_0$ and $B=0$ corresponds to the BCS results:
$b_{\mathrm{BCS}}(T_0) = N_0 \frac{7 \zeta(3)}{16 (\pi T_0)^2}$.
$Q_0 = 2n$ is the superfluid density at $T=0$ and $B=0$.
$\tilde{Q}_0 = 2n/\Delta_0^2$ represents $\tilde{Q}$ at $T=0$ and $B=0$.
The characteristic properties shown in Fig.~\ref{fig:delta_q_b_zeeman}
almost identical to those in Fig.~\ref{fig:delta_q_b_j32}.

In Table~\ref{table1}, we summarize 
the obtained results for three theoretical models of superconducting state.
The second row shows the structure of the pair potentials.
The third row shows the perturbations that generate the odd-frequency pairing correlations.
Since
$f^{\mathrm{odd}} \propto i \omega_n
[ H_{\mathrm{N}}(\bm{k}) \, \Delta(\bm{k}) - \Delta(\bm{k}) \, \undertilde{H}_{\mathrm{N}} (\bm{k}) ]$,
it is easy to confirm the presence of
odd-frequency pairs when the normal state Hamiltonian 
$H_{\mathrm{N}}$ includes the perturbations on the table. 
Although these three models describe different superconducting states
in different electronic structures, 
the coefficient $b$ and the superfluid density $Q$
share essentially the same expression as shown in Eq.~\eqref{eq:bq-relation}.
The existence of odd-frequency Cooper pairs is a common feature among the three 
uniform superconducting states.
The discontinuous transition occurs because odd-frequency Cooper pairs decrease the superfluid 
density at finite temperature.
The results displayed on Table~\ref{table1} suggest the universality of the phenomenon.
In this paper, we find a sufficient condition that makes the transition to the superconducting phase 
discontinuous.

\subsection{A relating state}

Fulde-Ferrell-Larkin-Ovchinnikov (FFLO) state is expected at high Zeeman fields~\cite{fulde_fflo,larkin_fflo}.
For $\mu_{\mathrm{B}}B / 2 \pi T_0 \gtrsim 0.18$, such spatially 
oscillating states can be a stable solution of the Gor'kov equation~\cite{matsuda:jpsj2007}.
Theoretical studies~\cite{burkhart:annphys1994,matsuo:jpsj1998} showed 
that the transition from the normal state to the FFLO state
can be both first and second-order depending on model parameters.
Since odd-frequency Cooper pairs also exist in such regime~\cite{chakraborty:prb2022,shu:prb2023},
there might be nontrivial relationships between the non-uniform odd-frequency Cooper pairs and the order of the phase transition.
However, the problem is beyond the scope of this paper and is left for our future study.

\section{Conclusion}
\label{sec:conclusion}
We have theoretically studied the thermodynamic instability 
of uniform superconducting states that include 
the subdominant pairing correlations belonging to
the odd-frequency symmetry class.
We especially focus on roles of odd-frequency Cooper pairs in the discontinuous transition  
to the superconducting phase.
In $j = 3/2$ superconductors,
we analyzed the contributions of the odd-frequency 
pairing correlations to 
the coefficient of $\Delta^4$ term in the Ginzburg-Landau (GL) free-energy $b$ and the superfluid density $Q$.
The odd-frequency pairing correlations decrease $b$ and $Q$ 
down to zero in the same manner because odd-frequency Cooper pairs are paramagnetic.
Since the effects are considerable at finite temperatures, the transition to 
a superconducting phase becomes discontinuous.
At a low temperature far below the transition temperature, on the other hand,
the pair potential and the superfluid density remain unchanged even in the presence of 
odd-frequency pairs.
The dependence of the odd-frequency pairing correlation functions on the Matsubara frequency 
explains well such characteristic features of the instability.

We also analyze the discontinuous transition in other superconductors such as
 a conventional $s$-wave spin-singlet superconductor under Zeeman fields and
 a two-band superconductor with interband pairing order.
 If odd-frequency Cooper pairs significantly reduce the superfluid density, 
 the transitions to these superconducting states can be discontinuous.
The coefficient $b$ and the superfluid density $Q$ calculated for these states also
share essentially the same expressions.
We conclude that the discontinuous transition to the uniform superconducting state
is a common feature of superconductors
in which the amplitude of the odd-frequency pairing correlation function
is sufficiently large.

\section*{Acknowledgments}
The authors are grateful to A.~A.~Golubov, Y.~Tanaka, Ya.~V.~Fominov and S.~Hoshino 
for useful discussions.
T. S. was supported by JST, the establishment of university fellowships towards the creation of science technology innovation, Grant Number JPMJFS2101.
S. K. was supported by JSPS KAKENHI (Grants No. JP19K14612 and No. JP22K03478) and JST 
CREST (Grant No. JPMJCR19T2).

\appendix

\section{Algebras of $\gamma$ matrices}
\label{sec:algebras}
The dispersions in the normal state Hamiltonian
are given by
%
%
\begin{align}
    \xi_{\bm{k}} &= \left( -2 t_1 - \frac{5}{2} t_2 \right) \sum_{\nu} \cos{k_{\nu}} + 6 t_1 + \frac{15}{2} t_2 - \mu , \\
    \epsilon_{\bm{k},1} &= 4\sqrt{3} t_3 \sin{k_x} \sin{k_y} , \\
    \epsilon_{\bm{k},2} &= 4\sqrt{3} t_3 \sin{k_y} \sin{k_z} , \\
    \epsilon_{\bm{k},3} &= 4\sqrt{3} t_3 \sin{k_z} \sin{k_x} , \\
    \epsilon_{\bm{k},4} &= \sqrt{3} t_2 ( -\cos{k_x} + \cos{k_y} ) , \\
    \epsilon_{\bm{k},5} &= t_2 ( -2 \cos{k_z} + \cos{k_x} + \cos{k_y} ).
    \label{eq:hn_gamma}
\end{align}
The spinors for the angular momentum of $j=3/2$ are described by,
\begin{align}
J_x &= \frac{1}{2}\left[\begin{array}{cccc}
0 & \sqrt{3} & 0 & 0 \\
\sqrt{3} & 0 & 2 & 0\\
0 & 2 & 0 & \sqrt{3} \\
0 & 0 & \sqrt{3} & 0
\end{array}\right], \\
J_y &= \frac{1}{2}\left[\begin{array}{cccc}
0 & -i\sqrt{3} & 0 & 0 \\
i\sqrt{3} & 0 & -2i & 0\\
0 & 2i & 0 & -i\sqrt{3} \\
0 & 0 & i\sqrt{3} & 0
\end{array}\right], \\
J_z &= \frac{1}{2}\left[\begin{array}{cccc}
3 & 0 & 0 & 0 \\
0 & 1 & 0 & 0\\
0 & 0 & -1 & 0 \\
0 & 0 & 0 & -3
\end{array}\right].
\end{align}
The five Dirac's $\gamma$-matrices are defined in $4 \times 4$ pseudospin space as
\begin{align}
\gamma^1=& \frac{1}{\sqrt{3}} (J_x J_y +J_y J_x), \;
\gamma^2= \frac{1}{\sqrt{3}} (J_y J_z +J_z J_y), \\
\gamma^3=& \frac{1}{\sqrt{3}} (J_z J_x +J_x J_z), \;
\gamma^4= \frac{1}{\sqrt{3}} (J_x^2- J_y^2), \\
\gamma^5=& \frac{1}{3} (2J_z^2 - J_x^2 -J_y^2), \label{eq:gamma5}
\end{align}
and $1_{4\times 4}$ is the identity matrix.
They satisfy the following relations
\begin{align}
    &\gamma^\mathrm{i}\, \gamma^\mathrm{j} + \gamma^\mathrm{j}\, \gamma^\mathrm{i} =2 \times  1_{4\times 4} \delta_{\mathrm{i}, \mathrm{j}}, \\
    &\gamma^1\, \gamma^2\, \gamma^3\, \gamma^4\, \gamma^5=-1_{4\times 4},\\ 
    &\{\gamma^\mathrm{i}\}^\ast = \{\gamma^\mathrm{i}\}^{\mathrm{T}}= U_T\, \gamma^\mathrm{i} \, U_T^{-1},\quad U_T=\gamma^1\, \gamma^2,
\end{align}
where $U_T$ is the unitary part of the 
time-reversal operation $\mathcal{T}=U_T \, \mathcal{K}$ with $\mathcal{K}$ meaning complex conjugation.
Eq.~\eqref{eq:hn} corresponds to the Luttinger-Kohn Hamiltonian~\cite{luttinger:pr1955} with cubic anisotropy
when we expand the trigonometric functions up to the second order of the momentum.

\section{Pseudospin-singlet pairing order}
\label{sec:singlet_j=3/2}
Pseudospin-singlet pair potential in the $j=3/2$ model is described by
\begin{align}
    \Delta (\bm{k}) &= \Delta \, U_T ,
\end{align}
in the BdG Hamiltonian in Eq.~\eqref{eq:Hbdg_j3/2}~~\cite{agterberg:prl2017,brydon:prb2018}.
Here, $\Delta$ is chosen to be real.
The anomalous Green's function within the first order of $\Delta$ results in,
\begin{align}
    \label{eq:f1_singlet}
    \mathcal{F}_1^{\mathrm{singlet}} &=
    \frac{\Delta}{Z_0} \left[
        -( \omega_n^2 + \xi_{\bm{k}}^2 + \vec{\epsilon}_{\bm{k}}^{\,2} ) + 2\xi_{\bm{k}} \vec{\epsilon}_{\bm{k}} \cdot \vec{\gamma}
    \right] U_T .
\end{align}
In this pairing order, the spin-orbit interaction does not induce any 
odd-frequency pairing correlations
but generates an even-frequency pseudospin-quintet pairing correlation described by the second term in Eq.~\eqref{eq:f1_singlet}.
The coefficients in the GL free-energy functional are expressed as
\begin{align}
    \label{eq:gla_singlet}
    a^{\mathrm{singlet}} &=
    \frac{1}{\tilde{g}} + T\sum_{\omega_n} \frac{1}{N}\sum_{\bm{k}}
    \frac{-2}{Z_0} ( \omega_n^2 + \xi_{\bm{k}}^2 + \vec{\epsilon}_{\bm{k}}^{\,2} ) , \\
    \label{eq:glb_singlet}
    b^{\mathrm{singlet}} &=
    T\sum_{\omega_n} \frac{1}{N}\sum_{\bm{k}}
    \frac{1}{Z_0^2}
    \left\{
        ( \omega_n^2 + \xi_{\bm{k}}^2 + \vec{\epsilon}_{\bm{k}}^{\,2} )^2
        + 4 \xi_{\bm{k}}^2 \vec{\epsilon}_{\bm{k}}^{\,2}
    \right\} ,
\end{align}
where $a^{\mathrm{singlet}}$ and $b^{\mathrm{singlet}}$ represent second and fourth-order coefficients of the GL functional, respectively.
The expression of $a^{\mathrm{singlet}}$ is equivalent to $a$ in Eq.~\eqref{eq:a_w/o_odd}
and $b^{\mathrm{singlet}} > 0$ holds true.
Therefore, the thermal property of the pseudospin-singlet state is identical to that of the BCS state
as well as the pseudospin-quintet states without odd-frequency Cooper pairs discussed in Sec.~\ref{subsec:gl_expansion}.

\section{Paramagnetic property of odd-frequency Cooper pair}
\label{sec:paraodd}
We consider a general Bogoliubov-de Gennes Hamiltonian describing electronic states of a uniform superconductor:
\begin{align}
    H(\bm{k}) &=
    \left[
        \begin{array}{cc}
            H_{\mathrm{N}}(\bm{k}) & \Delta(\bm{k}) \\
            -\undertilde{\Delta}(\bm{k}) & -\undertilde{H}_{\mathrm{N}}(\bm{k})
        \end{array} 
    \right],
    \label{eq:Hbdg}
\end{align}
where $\undertilde{X}(\bm{k},i\omega_n) = X^{\ast}(-\bm{k},i\omega_n)$ represents particle-hole conjugation.
We assume $H(\bm{k})$ is a $2M \times 2M$ matrix with $M$ being a positive integer.
$H(\bm{k})$ has particle-hole symmetry described as
\begin{align}
    \label{eq:PHS}
    C H(-\bm{k}) C^{-1} = - H(\bm{k}) , \quad
    C = \tau_1 \mathcal{K} ,
\end{align}
where $C$ represents charge-conjugation operator and $\tau_j$ for $j=1-3$ are Pauli matrices in the particle-hole space.
When we examine a response of superconductors to external perturbations within the linear response theory,
we often need to compute a correlation function of this form:
\begin{align}
    \label{eq:Pi}
    \Pi &= T\sum_{\omega_n} \frac{1}{V_{\mathrm{vol}}} \sum_{\bm{k}} A^2 \mathrm{Tr}
    \left[ \mathcal{G} \mathcal{G} - \undertilde{\mathcal{F}} \mathcal{F} \right]_{(\bm{k},i\omega_n)} ,
\end{align}
where $A$ represents a vertex in a corresponding correlation function
(e.g., current operator in a current-current correlation function)
and $\mathcal{G}$ $(\mathcal{F})$ is normal (anomalous) Green's function.
The Green's function is calculated by
the Gor'kov equation,
\begin{align}
    \left[ i\omega_n - H(\bm{k}) \right]
    \left[
        \begin{array}{rr}
            \mathcal{G} & \mathcal{F} \\
            -\undertilde{\mathcal{F}} & -\undertilde{\mathcal{G}}
        \end{array}
    \right]_{(\bm{k},i\omega_n)}
    = 1 .
\end{align}
The relation
%
\begin{align}
    \mathcal{F}^{\mathrm{T}}(-\bm{k},-i\omega_n) &= -\mathcal{F}(\bm{k},i\omega_n) , \label{eq:antisym}
\end{align}
%
holds true by  the Fermi-Dirac statistics of electrons.
The contribution of the anomalous Green's function to the correlation function is calculated to be
\begin{align}
    \Pi^{\mathcal{F}}
    &= T\sum_{\omega_n} \frac{1}{V_{\mathrm{vol}}} \sum_{\bm{k}} A^2 \mathrm{Tr}
    \left[ -\undertilde{\mathcal{F}}(\bm{k},i\omega_n) \mathcal{F}(\bm{k},i\omega_n) \right] \nonumber \\
    &=  T\sum_{\omega_n} \frac{1}{V_{\mathrm{vol}}} \sum_{\bm{k}} A^2 \sum_{\alpha \beta}
        \mathcal{F}_{\beta \alpha}^{\ast}{(\bm{k},-i\omega_n)} \mathcal{F}_{\beta \alpha}{(\bm{k},i\omega_n)} \nonumber \\
    &= T\sum_{\omega_n} \frac{1}{V_{\mathrm{vol}}} \sum_{\bm{k}} \nonumber \\
    &\hspace{1em}\times A^2 \sum_{\alpha \beta}
    \left(
        |f^{\mathrm{e}}_{\beta \alpha} (\bm{k},i\omega_n)|^2 - |f^{\mathrm{o}}_{\beta \alpha} (\bm{k},i\omega_n)|^2
    \right) ,
    \label{eq:Pif}
\end{align}
%
where we used the relation in Eq.~\eqref{eq:antisym} to reach the second line.
$\mathcal{F}_{\beta \alpha}$ represents $(\beta, \alpha)$-component of the $M \times M$ matrix $\mathcal{F}$ and
\begin{align}
    f^{\mathrm{e/o}}_{\beta \alpha}(\bm{k},i\omega_n) =
    \frac{\mathcal{F}_{\beta \alpha}(\bm{k},i\omega_n) \pm \mathcal{F}_{\beta \alpha}(\bm{k},-i\omega_n)}{2} ,
\end{align}
represents even-/odd-frequency components of $\mathcal{F}_{\beta \alpha}$.
Eq.~\eqref{eq:Pif} clearly shows the anomalous properties of odd-frequency Cooper pairs
by the negative sign of the second term.
When we consider a linear response to a static transverse vector potential,
the second term indicates that odd-frequency pairing correlations always have negative contributions to the Meissner kernel.
In other words, odd-frequency Cooper pairs exhibit a paramagnetic response to external magnetic fields
and then destabilize superconductivity by disturbing phase coherence.
Actually, above arguments are not valid when the corresponding vertex can not be factorized like Eq.~\eqref{eq:Pi}.
It has been shown that diamagnetic odd-frequency Cooper pairs can exist in some special systems~\cite{schmidt:prb2020,parhizgar:prb2021}.
But the odd-frequency Cooper pairs in most multiband/orbital superconductors show paramagnetic response~\cite{asano:prb2015}
and those considered in this paper are also paramagnetic.

\section{Induced pairing correlations}
\label{sec:triplet-septet}

The odd-frequency pairing correlations induced by the spin-orbit interactions are represented 
by $f_1^{\mathrm{odd}}(\boldsymbol{k}, i\omega_n) = i \omega_n\, P_{\mathrm{O}}\, U_T$.
We supply the calculated results of $P_{\mathrm{O}}$ for $(\eta_{\boldsymbol{k},4}, \eta_{\boldsymbol{k},5})
=(0, 1), (1,1)/\sqrt{2}$ and $(1,i)/\sqrt{2}$, 
\begin{align}
    P_{\mathrm{O}}^{(0,1)}
    &= 2 \gamma^5 \sum_{\mathrm{i} \neq 5} \epsilon_{\bm{k}, \mathrm{i}} \gamma^{\mathrm{i}} , \\
    P_{\mathrm{O}}^{(1,1)/\sqrt{2}}
    &= \sqrt{2} \bigg[
        ( \gamma^4 + \gamma^5 ) \sum_{\mathrm{i}=1}^{3}
        \epsilon_{\bm{k}, \mathrm{i}} \gamma^{\mathrm{i}} \nonumber\\
    &\hspace{3em}  + ( \epsilon_{\bm{k}, 5} - \epsilon_{\bm{k}, 4} ) \gamma^4 \gamma^5 \bigg] , \\
    P_{\mathrm{O}}^{(1,i)/\sqrt{2}}
    &= \sqrt{2} \bigg[
        ( \gamma^4 + i\gamma^5 ) \sum_{\mathrm{i}=1}^{3}
        \epsilon_{\bm{k}, \mathrm{i}} \gamma^{\mathrm{i}} \nonumber\\
    &\hspace{3em} + ( \epsilon_{\bm{k}, 5} - i\epsilon_{\bm{k}, 4} ) \gamma^4 \gamma^5
    \bigg] .
\end{align}
The expression of the coefficient $b$ for $(1,i)/\sqrt{2}$ state
is given by
\begin{align}
    b^{\mathrm{TRSB}} &=
    \label{eq:btrsb_gen}
    T\sum_{\omega_n} \frac{1}{N}\sum_{\bm{k}}
    \frac{1}{Z_0^2}
    \left[
        2\left( \omega_n^2 + \xi_{\bm{k}}^2 - A_0 \right. \right. \nonumber \\
        & \left.+ |\vec{\epsilon}_{\bm{k}} \cdot \vec{\eta}_{\bm{k}}|^2 \right)^2
    \left. -8 \omega_n^2 ( A_0 - |\vec{\epsilon}_{\bm{k}} \cdot \vec{\eta}_{\bm{k}}|^2 )
    \right] .
\end{align}

\section{Sixth-order coefficient}
\label{sec:sixth_coef}
The sixth-order coefficient of the GL functional in Eq.~\eqref{eq:gl_func} is represented as
\begin{align}
	c \Delta^6 &= T\sum_{\omega_n} \frac{1}{N}\sum_{\bm{k}}
    \frac{1}{6} \mathrm{Tr} \left[ \mathcal{F}_1 (\bm{k},i\omega_n) \Delta^{\dag}(\bm{k}) \right.\nonumber\\
 & \times   \left.  \mathcal{F}_1 (\bm{k},i\omega_n) \Delta^{\dag}(\bm{k}) \mathcal{F}_1 (\bm{k},i\omega_n) \Delta^{\dag}(\bm{k}) \right] .
\end{align}
When we choose $\left( \eta_{\bm{k},4},\eta_{\bm{k},5} \right) = (1,0)$ and $t_2=0$ in Eq.~\eqref{eq:Hbdg_j3/2}
as we assumed in Sec.~\ref{sec:discontinuous},
the sixth-order coefficient of the GL functional results in,
\begin{align}
    c = T\sum_{\omega_n} \frac{1}{N}\sum_{\bm{k}}
    \frac{-2}{3Z_0^3}
    &\left\{ ( \omega_n^2 + \xi_{\bm{k}}^2 - \vec{\epsilon}_{\bm{k}}^{\,2} )^3 \right. \nonumber \\
    &\left. - 12\omega_n^2 \vec{\epsilon}_{\bm{k}}^{\,2} ( \omega_n^2 + \xi_{\bm{k}}^2 - \vec{\epsilon}_{\bm{k}}^{\,2} ) \right\} .
    \label{eq:glc}
\end{align}
The first term in Eq.~\eqref{eq:glc} originates from the even-frequency correlation function.
On the other hand, the second term is composed of both even and odd-frequency correlation function.
Although eighth-order coefficients and above are also modified by odd-frequency correlation functions,
it is difficult to extract the physical meaning from these coefficients due to the cross terms.

\begin{widetext}
\section{Linear response to electromagnetic fields in a lattice model}
\label{sec:em_response}
The coupling between an electron and an electromagnetic field
is considered through the Peierls phase
in the kinetic energy~\cite{peierls:zphys1933,luttinger:pr1951}:
\begin{align}
    \mathcal{H}^{\mathrm{kin}}_{\mathrm{N}} &=
    -t_1 \sum_{j_z} \sum_{\langle \bm{r}_{\mathrm{i}}, \bm{r}_{\mathrm{j}} \rangle}
    e^{i \varphi_{\mathrm{ij}}}
    c^{\dag}_{\bm{r}_{\mathrm{i}},j_z} c_{\bm{r}_{\mathrm{j}},j_z} + \mathrm{H.c.} , \quad
    \varphi_{\mathrm{ij}} =
    e \int_{\bm{r}_\mathrm{j}}^{\bm{r}_\mathrm{i}} d \bm{r} \cdot \bm{A} (\bm{r}) ,
\end{align}
where $c^{\dag}_{\bm{r},j_z}$ ($c_{\bm{r},j_z}$) is the creation (annihilation) operator
for the electron at $\bm{r}$ with pseudospin $j_z$. 
We neglect the correction to the weak spin-orbit interactions ($t_3 \ll t_1$).
The current density operator $\bm{j}$ is defined from the variation of the Hamiltonian
with respect to the vector potential:
\begin{align}
    \delta \mathcal{H} (t) &= -\sum_{\bm{r}} \bm{j} (\bm{r},t) \cdot \delta \bm{A} (\bm{r},t) .
\end{align}
Within the first order of the vector potential, the current can be decomposed 
into the paramagnetic and diamagnetic terms,
\begin{align}
    j_{\mu} (\bm{r},t) &= j^{\mathrm{para}}_{\mu} (\bm{r}) + j^{\mathrm{dia}}_{\mu} (\bm{r},t) \quad
    (\mu = x, \, y, \, z) , \\
    j^{\mathrm{para}}_{\mu} (\bm{r}) &=
    iet_1 \sum_{j_z}
    \left[
        c^{\dag}_{\bm{r}+\hat{\bm{r}}_{\mu},j_z} c_{\bm{r},j_z} - \mathrm{H.c.}
    \right] , \quad
    j^{\mathrm{dia}}_{\mu} (\bm{r},t) =
    e^2 k_{\mu}(\bm{r}) A_{\mu} (\bm{r},t) ,
\end{align}
where $\hat{\bm{r}}_{\mu}$ is the basic lattice vector along the $\mu$-direction of a simple cubic lattice
and $k_{\mu}(\bm{r})$ is local kinetic energy operator with respect to the $\mu$-oriented links, which is defined as
\begin{align}
    k_{\mu} (\bm{r}) &=
    -t_1 \sum_{j_z}
    \left[
        c^{\dag}_{\bm{r}+\hat{\bm{r}}_{\mu},j_z} c_{\bm{r},j_z} + \mathrm{H.c.}
    \right] .
\end{align}
The perturbation Hamiltonian $\mathcal{H}'$ within the first order of the vector potential reads,
\begin{align}
    \mathcal{H}'(t) &= -\sum_{\bm{r}} \bm{j}^{\mathrm{para}} (\bm{r}) \cdot \bm{A} (\bm{r},t) .
\end{align}
In Sec.~\ref{sec:pairdens}, we examine the linear response in the $x$-direction:
\begin{align}
    j_{x} (\bm{q},\omega) &= - K_{xx} (\bm{q},\omega) A_x (\bm{q},\omega) .
\end{align}
The response kernel $K_{xx}$ is calculated to be~\cite{scalapino:prl1992,scalapino:prb1993,kostyrko:prb1994}
\begin{align}
    K_{xx} (\bm{q},\omega) &=
    e^2 \Braket{-k_x (\bm{r})} - \Lambda^{\mathrm{R}}_{xx} (\bm{q},\omega) ,
\end{align}
where $\Braket{-k_x (\bm{r})}$ represents the kinetic energy along the $x$-direction per unit cell and
$\Lambda^{\mathrm{R}}_{xx}$ is the current-current correlation function expressed as
\begin{align}
    \Lambda^{\mathrm{R}}_{xx} (\bm{q},\omega) &= \Lambda_{xx} (\bm{q},i\nu_m \rightarrow \omega + i\delta) , \\
    \Lambda_{xx} (\bm{q},i\nu_m) &=
    -e^2 T\sum_{\omega_n} \frac{1}{N} \sum_{\bm{k}}
    4t_1^2 \sin^2 \left( k_x + \frac{q_x}{2} \right) \nonumber \\
    &\times \mathrm{Tr}
    \left[ \mathcal{G} (\bm{k} + \bm{q}, i\omega_n + i\nu_m) \mathcal{G} (\bm{k}, i\omega_n) 
    -\undertilde{\mathcal{F}} (\bm{k} + \bm{q}, i\omega_n + i\nu_m) \mathcal{F} (\bm{k}, i\omega_n) \right] ,
\end{align}
where $\nu_m = 2m \pi T$ is a bosonic Matsubara frequency with $m$ being an integer
and $\delta$ is a small positive real value.
The superfluid density is defined by
\begin{align}
Q=\frac{K_{xx}(\boldsymbol{q}\to 0, \omega =0)}{2e^2 t_1}.
\end{align}
The contribution of odd-frequency pairing correlations in Sec.~\ref{sec:pairdens} is described by using
\begin{align}
Q^{\mathcal{F}}=     \
     T\sum_{\omega_n} \frac{1}{N} \sum_{\bm{k}}
    2t_1 \sin^2 k_x
 \mathrm{Tr}
    \left[ 
    -\undertilde{\mathcal{F}} (\bm{k}, i\omega_n) \mathcal{F} (\bm{k}, i\omega_n) \right]. 
\end{align}
The summation over Matsubara frequencies can be carried out analytically, 
when we use the spectral representation of the Green's function,
\begin{align}
    G(\bm{k},i\omega_n) &=
    \sum_{\lambda} \frac{\vec{\phi}_{\bm{k},\lambda} \vec{\phi}^{\,\dag}_{\bm{k},\lambda}}{i\omega_n-E_{\lambda}(\bm{k})} .
\end{align}
Here, the summation is taken over all eight indices of the eigenstates of the BdG Hamiltonian
and $\vec{\phi}_{\bm{k},\lambda}$
is the eigenvector belonging to the eigenenergy $E_{\lambda}(\bm{k})$.
After the summation over the Matsubara frequencies, we reach at  
\begin{align}
    \Braket{-k_x (\bm{r})} &=
    \frac{1}{N} \sum_{\bm{k},\lambda=\mathrm{S}\pm} 2t_1 \cos k_x 
    \left[
        u_{\bm{k}}^2 f(E_{\lambda}) + v_{\bm{k}}^2 f(-E_{\lambda})
    \right] ,
\end{align}
with
\begin{align}
    u_{\bm{k}}^2 = 1 + \frac{\xi_{\bm{k}}}{\sqrt{\xi_{\bm{k}}^2 + \Delta^2}} , \quad
    v_{\bm{k}}^2  = 1 - \frac{\xi_{\bm{k}}}{\sqrt{\xi_{\bm{k}}^2 + \Delta^2}} , \quad
\end{align}
and
\begin{align}
    \Lambda^{\mathrm{R}}_{xx} (\boldsymbol{q} \to 0, \omega=0) = e^2 \frac{1}{N} \sum_{\bm{k},\lambda=\mathrm{S}\pm} 8t_1^2 \sin^2 k_x
    \left( -\frac{\partial f(E_{\lambda})}{\partial E_{\lambda}} \right) ,
\end{align}
for $(\eta_{\bm{k}, 4},\eta_{\bm{k}, 5})=(1,0)$, $(0,1)$, and $(1,1)/\sqrt{2}$ at $t_2=0$.

The Green's function for $(\eta_{\bm{k}, 4},\eta_{\bm{k}, 5})=(1,0)$, $(0,1)$, and $(1,1)/\sqrt{2}$
at $t_2=0$ is calculated to be
\begin{align}
    \mathcal{G}(\bm{k},i\omega_n) &=
    -\frac{1}{Z}
    \left[
        (\omega_n^2 + \xi_{\bm{k}}^2  + \Delta^2) (i\omega_n + \xi_{\bm{k}}) 
		+\vec{\epsilon}_{\bm{k}}^{\,2} (i\omega_n - \xi_{\bm{k}})  
   -     \left\{ (i\omega_n + \xi_{\bm{k}})^2 + \Delta^2 - \vec{\epsilon}_{\bm{k}}^{\,2}
		\right\} \vec{\epsilon}_{\bm{k}} \cdot \vec{\gamma}
    \right] , \label{eq:g-func} \\
    \mathcal{F}(\bm{k},i\omega_n) &=
    -\frac{\Delta}{Z}
    \left[
        \omega_n^2 + \xi_{\bm{k}}^2  + \Delta^2 -\vec{\epsilon}_{\bm{k}}^{\,2} -2i\omega_n \vec{\epsilon}_{\bm{k}} \cdot \vec{\gamma}
    \right]
    \vec{\eta}_{\bm{k}} \cdot \vec{\gamma} U_{T} , \label{eq:f-func}\\
	Z &=(\omega_n^2 + \xi_{\bm{k}}^2 + \Delta^2- \vec{\epsilon}_{\bm{k}}^{\,2}  )^2 + 4\omega_n^2 \vec{\epsilon}_{\bm{k}}^{\,2}. 
\end{align}
The last term in Eq~\eqref{eq:f-func} represents the odd-frequency pairing correlation
induced by the spin-orbit interaction.

\section{Discontinuous transition in a two-band superconductor with interband pairing order}
\label{sec:interband}
We consider following mean-field Hamiltonian which describes the two-band superconducting 
states with the interband pairing order,
\begin{align}
    \mathcal{H} &= \sum_{\bm{k}}
    \left[
        a_{\bm{k},\uparrow}^{\dag} \  a_{\bm{k},\downarrow}^{\dag} \  b_{\bm{k},\uparrow}^{\dag} \  b_{\bm{k},\downarrow}^{\dag}
        a_{-\bm{k},\uparrow} \  a_{-\bm{k},\downarrow} \  b_{-\bm{k},\uparrow} \  b_{-\bm{k},\downarrow}
    \right] \nonumber \\
    &\hspace{1em}\times
    \left[
        \begin{array}{cccc|cccc}
            \varepsilon_{\bm{k}}^{\mathrm{a}} & & V & & & & & \Delta \\
              & \varepsilon_{\bm{k}}^{\mathrm{a}} & & V & & & s\Delta & \\
            V & & \varepsilon_{\bm{k}}^{\mathrm{b}} & & & -s\Delta & & \\
              & V & & \varepsilon_{\bm{k}}^{\mathrm{b}} & -\Delta & & & \\ \hline
              & & & -\Delta & -\varepsilon_{\bm{k}}^{\mathrm{a}} & & -V & \\
              & & -s\Delta & & & -\varepsilon_{\bm{k}}^{\mathrm{a}} & & -V \\
              & s\Delta & & & -V & & -\varepsilon_{\bm{k}}^{\mathrm{b}} & \\
            \Delta & & & & & -V & & -\varepsilon_{\bm{k}}^{\mathrm{b}}
        \end{array}
    \right]
    \left[
        \begin{array}{c}
            a_{\bm{k},\uparrow} \\ a_{\bm{k},\downarrow} \\ b_{\bm{k},\uparrow} \\ b_{\bm{k},\downarrow} \\
            a_{-\bm{k},\uparrow}^{\dag} \\ a_{-\bm{k},\downarrow}^{\dag} \\ b_{-\bm{k},\uparrow}^{\dag} \\ b_{-\bm{k},\downarrow}^{\dag}
        \end{array}
    \right] ,
    \label{eq:H_inter}
\end{align}
where $a^{\dag}_{k\sigma}$ ($b^{\dag}_{k\sigma}$) is a creation operator of an electron
in band a (b) with momentum $\bm{k}$ and spin $\sigma \, (= \uparrow, \downarrow)$.
$\varepsilon_{\bm{k}}^{\mathrm{l}}$ ($\mathrm{l} = \mathrm{a}, \mathrm{b}$)
is defined by $\varepsilon_{\bm{k}}^{\mathrm{l}} = \frac{\bm{k}^2}{2m_{\mathrm{l}}} - \mu_{\mathrm{l}}$
and $V$ is the hybridization potential mixing the two bands.
$\Delta$ represents interband pairing potential belonging to $s$-wave spin-triplet odd-band-parity (spin-singlet even-band-parity) symmetry class 
when we choose $s=+1 \, (-1)$.
Eq.~\eqref{eq:H_inter} with $s=+1$ corresponds to the mean-field Hamiltonian considered in Ref.~\cite{silva:physletta2014}.
Eq.~\eqref{eq:H_inter} can be block-diagonalized and the reduced $4 \times 4$ Hamiltonian is represented by
\begin{align}
    \check{H}(\bm{k}) &=
    \left[
        \begin{array}{cc}
            \hat{H}_{\mathrm{N}} & \hat{\Delta} \\
            \hat{\Delta}^{\dag} & -\hat{H}_{\mathrm{N}} 
        \end{array}
    \right] , \quad
    \hat{H}_{\mathrm{N}} =
    \xi + \varepsilon \hat{\rho}_3 + V \hat{\rho}_1 , \quad
    \hat{\Delta} =
    \begin{cases}
        \Delta (i\hat{\rho}_2) & (s=+1) \\
        \Delta \hat{\rho}_1 & (s=-1)
    \end{cases} ,
    \label{eq:H_reduced}
\end{align}
where
$\xi         = (\varepsilon_{\bm{k}}^{\mathrm{a}} + \varepsilon_{\bm{k}}^{\mathrm{b}})/2$,
$\varepsilon = (\varepsilon_{\bm{k}}^{\mathrm{a}} - \varepsilon_{\bm{k}}^{\mathrm{b}})/2$,
and $\hat{\rho}_j$ for $j=1-3$ are Pauli matrices in the two-band space.
%
%
The anomalous Green's function $\hat{\mathcal{F}}$ is calculated as
\begin{align}
    \hat{\mathcal{F}}(\bm{k}, i\omega_n) &=
    \left\{
    \begin{array}{lc}
        \frac{-1}{Z_{+1}} \left[ \omega_n^2 + \xi^2 - \varepsilon^2 - V^2 + \Delta^2 - 2i\omega_n (\varepsilon \hat{\rho}_3 + V \hat{\rho}_1) \right]
        \Delta (i\hat{\rho}_2)
        & \quad (s=+1) \\
        \frac{-1}{Z_{-1}} \left[ \omega_n^2 + \xi^2 - \varepsilon^2 + V^2 + \Delta^2 - 2V\xi\hat{\rho}_1 + 2iV\varepsilon\hat{\rho}_2 -2i\omega_n \varepsilon\hat{\rho}_3 \right]
        \Delta \hat{\rho}_1
        & \quad (s=-1)
    \end{array}
    \right. ,
    \label{eq:f-func_inter} \\
    Z_{+1} &= (\omega_n^2 + \xi^2 - \varepsilon^2 - V^2 + \Delta^2)^2 + 4\omega_n^2 (\varepsilon^2 + V^2) , \\
    Z_{-1} &= (\omega_n^2 + \xi^2 - \varepsilon^2 + V^2 + \Delta^2)^2 - 4(V^2\xi^2 - V^2\varepsilon^2 - \omega_n^2\varepsilon^2) .
\end{align}
$- 2i\omega_n \varepsilon \hat{\rho}_3 \Delta (i\hat{\rho}_2)$ and
$- 2i\omega_n V \hat{\rho}_1 \Delta (i\hat{\rho}_2)$
in the numerator for $s=+1$ and
$- 2i\omega_n \varepsilon \hat{\rho}_3 \Delta \hat{\rho}_1$
in that for $s=-1$
represent
odd-frequency pairing correlations.

The authors of Ref.~\cite{silva:physletta2014} analyzed 
the transition from the normal state to the superconducting state
described by $\mathcal{H}$ in Eq.~\eqref{eq:H_inter} with $s=+1$.
They found that the transition
becomes first-order under the sufficiently large band hybridization $V$.
The mechanism is explained well by the paramagnetic property of the odd-frequency Cooper pairs induced by $V$
as well as we discussed in this paper.
Moreover, the discontinuous transition is also expected 
in the presence of sufficiently large asymmetry between the two bands $\varepsilon$.
In this case, the odd-frequency pairing correlation is induced also by the band asymmetry $\varepsilon$.
This argument is valid 
because $\check{H}(\bm{k})$ in Eq.~\eqref{eq:H_reduced} for $s=+1$ is equivalent to the Hamiltonian of a spin-singlet superconductor under Zeeman fields.
The calculated results of the fourth-order coefficient of the GL free-energy and the superfluid density
are given by
\begin{align}
    b \propto Y_{\mathrm{inter}}(A_0,C_0) , \quad
    Q \propto \Delta^2 Y_{\mathrm{inter}}(A,C) ,
    \label{eq:b_q_interband}
\end{align}
with $Y_{\mathrm{inter}} = Y|_{\mu_{\mathrm{B}}B \rightarrow \sqrt{\epsilon^2 + V^2}}$ in Eq.~\eqref{eq:yac}.
Here we consider a simple band structure $m_{\mathrm{a}}=m_{\mathrm{b}}$~\cite{asano:prb2015} for simplicity.

\section{A spin-singlet superconductor under spin-dependent potentials}
\label{sec:singlet}
We consider a spatially uniform spin-singlet $s$-wave superconducting state under spin-dependent potentials.
The Gor'kov equation reads,
\begin{align}
&\left[ i\omega_n  - \check{H}_{\mathrm{BdG}} \right] 
\left[ \begin{array}{rr}
\mathcal{G} & \mathcal{F} \\ - \undertilde{\mathcal{F}} & - \undertilde{\mathcal{G}}
\end{array}\right]_{(\boldsymbol{k}, i\omega_n )} = \check{1} , \quad
\check{H}_{\mathrm{BdG}} = \left[ \begin{array}{cc} \hat{H}_{\mathrm{N}} & \hat{\Delta} \\
- \undertilde{\hat{\Delta}} & - \undertilde{\hat{H}}_{\mathrm{N}} \end{array}\right].
\end{align}
The anomalous Green's function is represented as
\begin{align}
&\mathcal{F}(\boldsymbol{k}, i\omega_n )= \left[ \hat{\Delta}\, \undertilde{\hat{\Delta}}
- \omega_n^2 - \hat{\Delta}  \, \undertilde{\hat{H}}_{\mathrm{N}}\, \hat{\Delta}^{-1}\,  \hat{H}_{\mathrm{N}}
+ i\omega_n \, P \,  \hat{\Delta}^{-1}
\right]^{-1} \hat{\Delta} , \label{eq:f_general}\\
&P= ( \hat{\Delta}  \, \undertilde{\hat{H}}_{\mathrm{N}} -  \hat{H}_{\mathrm{N}}\, \hat{\Delta}) , \quad \hat{\Delta}= \Delta \, i \hat{\sigma}_2, \label{eq:p-general}
\end{align}
where $\hat{\sigma}_{\mathrm{j}}$ for $\mathrm{j}=1-3$ are Pauli matrices in the spin space.
The odd-frequency pairing correlations appear for $P\neq 0$~\cite{triola:annphys2020}.
%
%
%
The electric current $\boldsymbol{j}$ within the 
linear response to a vector potential $\boldsymbol{A}$ is described by~\cite{agd}
\begin{align}
\boldsymbol{j} &= - \frac{e^2}{mc} \, Q \, \boldsymbol{A} , \quad
Q =  n \, T\sum_{\omega_n} \int d \xi \,
\langle \mathrm{Tr} \left[
    \mathcal{G}\, \mathcal{G} - \undertilde{\mathcal{F}} \, \mathcal{F} - \mathcal{G}_{\mathrm{N}}\,
    \mathcal{G}_{\mathrm{N}}
\right]_{(\boldsymbol{k}, i\omega_n)} \rangle_{\mathrm{FS}} ,
\end{align}
where $n$ is the electron density per spin, 
$\langle \cdots \rangle_{\mathrm{FS}} \equiv \int \frac{d \Omega}{4\pi} \cdots$ is the Fermi surface average, and
 $Q$ is referred to as superfluid density.

Firstly, we consider the normal state Hamiltonian including an antisymmetric spin-orbit interaction,
\begin{align}
&\hat{H}_{\mathrm{N}}(\boldsymbol{k}) = \xi_{\boldsymbol{k}} - \boldsymbol{\alpha}_{\boldsymbol{k}} \cdot \hat{\boldsymbol{\sigma}} , \quad
\xi_{\boldsymbol{k}} = \frac{\hbar^2 \boldsymbol{k}^2}{2m} - \mu, \quad \boldsymbol{\alpha}_{-\boldsymbol{k}} = -\boldsymbol{\alpha}_{\boldsymbol{k}}. 
\end{align}
The Fermi surface is split into two due to the spin-orbit interaction in the normal state.
The Green's functions are calculated to be
\begin{align}
    \mathcal{G}(\boldsymbol{k}, i\omega_n ) =& -
    \frac{(i\omega_n+\xi_{\boldsymbol{k}})(\omega_n^2 + \xi_{\boldsymbol{k}}^2 +\Delta^2)
    +(i\omega_n-\xi_{\boldsymbol{k}}) \boldsymbol{\alpha}_{\boldsymbol{k}}^2  
    + \left\{ (i\omega_n+\xi_{\boldsymbol{k}})^2 - \boldsymbol{\alpha}_{\boldsymbol{k}}^2
    -\Delta^2\right\} \boldsymbol{\alpha}_{\boldsymbol{k}} \cdot \hat{\boldsymbol{\sigma}}}{\left\{\omega_n^2 + \xi_{\boldsymbol{k}}^2 +\Delta^2
    + \boldsymbol{\alpha}_{\boldsymbol{k}}^2 \right\}^2 - 4 \xi_{\boldsymbol{k}}^2\,\boldsymbol{\alpha}_{\boldsymbol{k}} ^2 } , \\
    \mathcal{F}(\boldsymbol{k}, i\omega_n ) =& -\frac{\omega_n^2 + \xi_{\boldsymbol{k}}^2 +\Delta^2 + \boldsymbol{\alpha}_{\boldsymbol{k}} ^2  
    + 2\, \xi_{\boldsymbol{k}}\, \boldsymbol{\alpha}_{\boldsymbol{k}} \cdot \hat{\boldsymbol{\sigma}}}
    {\left\{\omega_n^2 + \xi_{\boldsymbol{k}}^2 +\Delta^2 + \boldsymbol{\alpha}_{\boldsymbol{k}}^2 \right\}^2 - 4 \xi_{\boldsymbol{k}}^2\,\boldsymbol{\alpha}_{\boldsymbol{k}} ^2  }
    \Delta\, i \hat{\sigma}_2.
\end{align}
The last term in the numerator of $\mathcal{F}$ represents the spin-triplet odd-parity pairing correlation 
induced by the spin-orbit interaction.
Since $P=0$, the odd-frequency component is absent.
The gap equation becomes
\begin{align}
    \Delta &= g \, T\sum_{\omega_n} \frac{1}{V_{\mathrm{vol}}} \sum_{\boldsymbol{k}} 
    \frac{1}{2}\mathrm{Tr}\left[
    \mathcal{F}(\boldsymbol{k}, i\omega_n ) \, i\hat{\sigma}_2 \right]
    = g \, N_0 \pi T\sum_{\omega_n} \frac{\Delta}{\sqrt{\omega_n^2+\Delta^2}} ,
\end{align}
which is identical to that in the BCS theory, where $N_0$ is the density of states at the Fermi level per spin.
It is possible to show that the superfluid density and the coefficients of the GL free-energy are
identical to those in the BCS theory,
\begin{align}
    Q_{\mathrm{BCS}} &= 2n \pi T \sum_{\omega_n}\frac{\Delta^2}{(\omega_n^2+\Delta^2)^{3/2}} , \quad
    a_{\mathrm{BCS}} = \frac{1}{g} - N_0 \pi T \sum_{\omega_n}\frac{1}{|\omega_n|} , \quad
    b_{\mathrm{BCS}} = \frac{N_0 \pi}{4} T \sum_{\omega_n} \frac{1}{|\omega_n|^3} = N_0 \frac{7 \zeta(3)}{16 (\pi T)^2} \label{eq:glab_bcs} ,
\end{align}
where $\zeta(n)$ is Riemann zeta function.
Therefore, the spin-orbit interactions do not change any thermal properties of a spin-singlet superconductor~\cite{frigeri:prl2004} as we discussed in the introduction. 

Secondly, we consider the normal state Hamiltonian including the Zeeman potential, 
\begin{align}
\hat{H}_{\mathrm{N}}(\boldsymbol{k}) = \xi_{\boldsymbol{k}} - \mu_{\mathrm{B}} \boldsymbol{B} \cdot \hat{\boldsymbol{\sigma}},
\end{align}
where $\mu_{\mathrm{B}}$ is Bohr's magneton and $\boldsymbol{B}$ is a Zeeman field. 
The odd-frequency pairing correlation appears because
 $P = 2\, \mu_{\mathrm{B}} \, \boldsymbol{B} \cdot \hat{\boldsymbol{\sigma}}$ remains finite.
The Green's functions are calculated as
\begin{align}
    &\mathcal{G}(\boldsymbol{k}, i\omega_n) = \frac{-1}{Z_{\mathrm{z}}}\left[
    (i\omega_n + \xi_{\boldsymbol{k}}) (\omega_n^2+\xi_{\boldsymbol{k}}^2+\Delta^2) + 
    (i\omega_n - \xi_{\boldsymbol{k}}) \mu_{\mathrm{B}}^2 B^2
    + 
    \left\{  (i\omega_n + \xi_{\boldsymbol{k}})^2 +\Delta^2 -\mu_{\mathrm{B}}^2 B^2 \right\} \mu_{\mathrm{B}} \boldsymbol{B} \cdot \hat{\boldsymbol{\sigma}}
    \right],\\
    &\mathcal{F}(\boldsymbol{k}, i\omega_n) =
    \frac{-1}{Z_{\mathrm{z}}}\left[ \omega_n^2 + \xi_{\boldsymbol{k}}^2 + \Delta^2 - \mu_{\mathrm{B}}^2 B^2
    +2 i \omega_n \, \mu_{\mathrm{B}} \, \boldsymbol{B} \cdot \hat{\boldsymbol{\sigma}} \right]
    \Delta i \hat{\sigma}_2 , \label{eq:fba1} \\
    &Z_{\mathrm{z}} =
    \xi_{\boldsymbol{k}}^4 + 2\, \xi_{\boldsymbol{k}}^2\, A + C , \quad
    A = \Delta^2+\omega_n^2-\mu_{\mathrm{B}}^2 B^2 , \quad
     C = A^2 + 4 \omega_n^2\, \mu_{\mathrm{B}}^2 B^2 . 
\end{align}
The last term in $\mathcal{F}$ represents the pairing correlation belonging to odd-frequency 
spin-triplet $s$-wave symmetry class.
%
%
The gap equation becomes
\begin{align}
    \Delta = g \, N_0 \pi T\sum_{\omega_n} \frac{\Delta \sqrt{A + \sqrt{C}}}{\sqrt{2C}} .
    \label{eq:gapeq_zeeman}
\end{align}
The self-consistent pair potential $\Delta_{\mathrm{eq}}$ satisfies
Eq.~\eqref{eq:gapeq_zeeman} and minimizes the thermodynamic potential.
The coefficients in the free-energy result in
\begin{align}
    a &= \frac{1}{g} - N_0 \pi T \sum_{\omega_n} 
    \frac{\omega_n^2}{|\omega_n| (\omega_n^2 + \mu_{\mathrm{B}}^2 B^2)} , \label{eq:glab_zeeman} \quad
    b = \frac{\sqrt{2}}{4} \, N_0 \pi T \sum_{\omega_n}
    \frac{A_0^3+\sqrt{C_0}(A_0^2-2\omega_n^2 \mu_{\mathrm{B}}^2 B^2) }
    {[ C_0 (A_0+\sqrt{C_0})]^{3/2}} , 
\end{align}
with $A_0 = A|_{\Delta=0}$ and $C_0=C|_{\Delta=0}$.
The second term of the coefficient $a$ in Eq.~\eqref{eq:glab_zeeman} becomes smaller than that in Eq.~\eqref{eq:glab_bcs}, 
which leads to the suppression of $T_c$.
The last term of the coefficient $b$ in Eq.~\eqref{eq:glab_zeeman} is derived from the odd-frequency pairing correlation function
and decreases the coefficient $b$.
The superfluid density is calculated to be
\begin{align}
    Q= 2\sqrt{2} n \pi T \sum_{\omega_n}
    \frac{\Delta^2 \left\{ A^3+\sqrt{C}(A^2-2\omega_n^2 \mu_{\mathrm{B}}^2 B^2) \right\} }
    {[ C (A+\sqrt{C})]^{3/2}}.\label{eq:q_zeeman}
\end{align}
The comparison between the expression of the superfluid density in Eq.~\eqref{eq:q_zeeman} and 
that of the coefficient $b$ in Eq.~\eqref{eq:glab_zeeman} shows that
the odd-frequency pairing correlation decreases $Q$ and $b$ in exactly the same manner.

\end{widetext}
\twocolumngrid
\nocite{apsrev42Control}

%

\end{document}